\documentclass[%
 reprint,
 amsmath,
 amssymb,
 aps
]{revtex4-1}

\usepackage{graphicx}% Include figure files
\usepackage{dcolumn}% Align table columns on decimal point
\usepackage{bm}% bold math

\usepackage{mathtools}
\usepackage{float}
\usepackage{siunitx}
\usepackage{multirow}

\begin{document}

\title{CSR wake expressions with two bending magnets \\ and simulation results for the multi-turn ERL, CBETA }

\author{W.~Lou}
\email{wl528@cornell.edu}
\author{G.H.~Hoffstaetter}
\affiliation{CLASSE, Cornell University, Ithaca, NY 14853, USA
}

\date{\today}

\begin{abstract}
This paper consists of two main parts regarding Coherent Synchrotron Radiation (CSR). The first part extends the CSR theory of two particle interaction from a system of one bending magnet to two bending magnets, where the wake can leak from the first to the second. The new theory agrees well with the established simulation code Bmad. The second part of the paper presents the CSR simulation results on CBETA, the Cornell BNL Energy-Recovery-Linac (ERL) Test Accelerator ~\cite{cbetacdr}, in which the magnets are so close to each other and the new extended theory becomes important. CBETA is the first multi-turn ERL with Superconducting Radio Frequency (SRF) accelerating cavities and a Fixed Field Alternating gradient (FFA) beamline. Simulations show that CSR causes phase space dilution that becomes more significant as the bunch charge and the number of recirculation passes increase. Potential ways to mitigate the CSR effects, including adding vacuum chamber shielding and increasing bunch length, are being investigated.
\end{abstract}

\maketitle

\section{Introduction}
Synchrotron radiation occurs when an electron traverses a curved trajectory, and the radiation emitted can give energy kicks to the other electrons in the same bunch. While the high frequency components of the radiation spectrum tend to add up incoherently, the low frequency components, with wavelength on the order of the bunch length, can add coherently. These are termed incoherent and coherent synchrotron radiation respectively (ISR and CSR). While the total intensity for ISR scales linearly with the number of charged particles ($N_p$), it scales as $N_p^2$ for CSR. 

The most elementary CSR theory studies the interaction of two electrons passing through one bending magnet, and the CSR wake $w = d\mathcal{E}_{\text{CSR}}/cdt$ has been derived for four different cases (A,B,C,D) in \cite{Saldin}, depending on the location of the source and the observation points within or outside the magnet. The approximated wake expressions $W(s)=\int w(s-s') \lambda(s') ds'$ for an arbitrary longitudinal bunch distribution $\lambda(s)$ have been calculated for the four cases in \cite{Emma}. However, if a second magnet is located downstream not sufficiently far away for the exit wake from the first magnet to attenuate, the wake leaks into the second magnet, and the existing formulas cannot be applied. Therefore, our goal is to derive the wake expressions for a system with two magnets. The derivation and analysis are shown in the first half of this paper.

For an ERL which aims for high beam quality like CBETA, CSR can pose detrimental effects on the beam bunches, including energy loss, increase in energy spread, and potential the micro-bunching instability \cite{Tsai_2017}. It is therefore important to run CSR simulations for CBETA, and investigate potential ways for mitigation. These will be shown in the second half of this paper.

\section{Two particle interaction}

The Lienard-Wiechert formula describes the electric field seen by the front electron at point $P$ (as in Fig.~\ref{caseA} for example) at time $t$. This field is produced by the tail electron at point $P'$ at retarded time $t'$ \cite{Jackson}:
\begin{equation}
\textbf{E}(P) =  \frac{ke}{\gamma^2}
\frac{(\textbf{L}-L\beta\textbf{n}')}{(L-\textbf{L} \boldsymbol{\cdot} \beta\textbf{n}')^3} +\frac{ke}{c^2}\frac{ (\textbf{L} \times [(\textbf{L}-L\beta\textbf{n}') \times \textbf{a}'])}{(L-\textbf{L} \boldsymbol{\cdot} \beta\textbf{n}')^3},
\label{LW}
\end{equation}

in which $k=1/(4 \pi \epsilon_0)$, $e$ is the electron charge, and $c$ is the speed of light. $\textbf{L}$ is the vector pointing from $P'$ to $P$, and $L$ is its magnitude. The two electrons are assumed to have the same speed $v = \beta c$. $\textbf{n}'$ and $\textbf{a}'$ are respectively the unit velocity vector and the acceleration at point $P'$. By convention we refer to the term in Eq.\eqref{LW} that is proportional to $1/\gamma^2$ as the velocity field, and the term with $\textbf{a}'$ as the acceleration field \cite{Saldin}.

Let $L_s$ denote the path length from $P'$ and $P$ travelled by the electron. Also, let $s$ and $s'$ denote the longitudinal position of the front and tail electron at time $t$ with respect to the bunch center (Note: $s'$ is evaluated at the observation time $t$, $not$ the retarded time $t'$). Then the distance between the two particles at time $t$ can be expressed as \cite{Sagan}:
\begin{equation}
    \Delta \equiv s-s' = L_s - \beta L.
\end{equation}

As we will see, $\Delta$ is an important quantity in deriving the wake expressions. As $\Delta \rightarrow 0$, the velocity field has a singularity of order $1/\Delta^2$, which is dealt with by splitting $\textbf{E}(P)$ into two terms: 

\begin{equation}
\textbf{E}(P) = \textbf{E}_{\text{SC}}  + \textbf{E}_{\text{CSR}}. 
\end{equation}

The singularity is contained in the space charge term:
\begin{equation}
\textbf{E}_{\text{SC}} = \frac{ke\textbf{n}}{\gamma^2\Delta^2},
\end{equation}
in which $\textbf{n}$ is the unit velocity vector at point $P$.
$\textbf{E}_{\text{SC}}$ is the field resulting from two particles moving on a straight line without acceleration. 

The rate of change in energy of the front electron at point $P$ is:
\begin{equation}
\frac{d\mathcal{E}}{dt} = \textbf{v} \boldsymbol{\cdot} \textbf{F} = ec\beta\textbf{n}\boldsymbol{\cdot}\textbf{E}(P).
\end{equation}

The CSR wake seen by the front electron is defined to be:
\begin{equation}
w \equiv \frac{d\mathcal{E}_{\text{CSR}}}{cdt} = e\beta\textbf{n}\boldsymbol{\cdot} (\textbf{E}(P) -\textbf{E}_{\text{SC}}).
\end{equation}

Our goal is to find $w$ for the cases with two bending magnets, and apply similar approximations in \cite{Emma} to solve for the wake expression $W(s)$ seen by the full bunch. Note that this is an one dimensional theory which assumes that all electrons move along the same path. Following the nomenclature in \cite{Saldin}, this will introduce four additional cases: E,F,G, and H. Case E and G are extension of case A and C in which $P'$ is located on the drift before the first magnet. We call these cases the ``odd cases". Similarly, case F and case H are extension of case B and D in which $P'$ is located within the first magnet, and we call them the ``even cases". In the two subsections below we will derive the wake expressions respectively for the four odd cases and four even cases. The main results for all eight cases are summarized in Appendix A.

\subsection{Odd cases}
We will first re-derive the wake expression for case A and C, then apply similar formulation to obtain the expressions for case E and G. 
\subsubsection{Case A}
\begin{figure}[!h]
	\centering
	\includegraphics[width=0.20\textwidth]{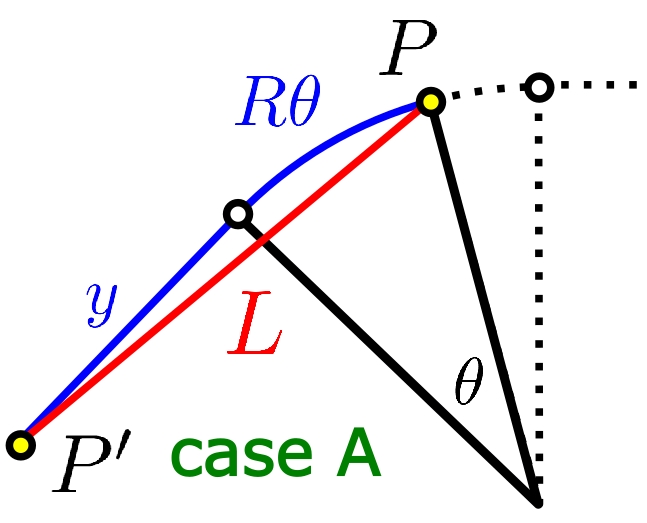}
	\caption{(Color) Geometry for case A. Point $P$ and $P'$ are labeled by yellow dots. The blue curve is the path traversed by electrons. These conventions apply to all other geometry figures in Section I.}
	\label{caseA}
\end{figure}

The geometry for case A is shown in Fig.~\ref{caseA}. The observation point $P$ is located at an angle $\theta$ into the magnet. Since $P'$ is located on a drift, i.e. $\textbf{a}' = 0$, as the acceleration field vanishes. Let us define 
\begin{gather}
\mathcal{N}_v \equiv  \beta\textbf{n} \boldsymbol{\cdot} (\textbf{L}-L\beta\textbf{n}'),\\
\mathcal{D} \equiv  (L-\textbf{L}\boldsymbol{\cdot}\beta\textbf{n}').
\end{gather}

The subscript $v$ indicates that the numerator belongs to the velocity field. Then we can write $w$ as:
\begin{eqnarray}
w = \frac{ke^2}{\gamma^2} \left(\frac{\mathcal{N}_v}{\mathcal{D}^3} - \frac{1}{\Delta^2}\right).
\end{eqnarray}
To find $\mathcal{N}_v$ and $\mathcal{D}$ we have to first fix our coordinate system. Let us pick the unit velocity vector at $P$ to be $\textbf{n} = \langle 1,0,0 \rangle$. Vector analysis shows that:
\begin{gather}
\textbf{n}' = \langle \cos{\theta},\sin{\theta},0 \rangle,\\
L_x=  y\cos{\theta} + L_c\cos{(\theta/2)},\\
L_y = y\sin{\theta} + L_c\sin{(\theta/2)},
\end{gather}

in which $L_c$ is the chord length (See Fig.(1)), and $\textbf{L} = \langle L_x, L_y, 0 \rangle$. As in \cite{Saldin}, we assume a small bending angle ($\theta \ll 1$), and expand all quantities, holding $R\theta$ constant, up to order of $\theta^2$. This yields:
\begin{gather}
L_c = R\theta - R\theta^3/24, \\
L = L_s - \frac{R\theta^3}{24}\frac{(R\theta+4y)}{L_s},  
\end{gather}
in which $L_s = (R\theta+y)$ for case A. 
Since $R\theta$ is held constant, terms like $R^n\theta^{(n+2)}$ do $not$ vanish $(n \in \mathbb{N})$. It follows that:
\begin{equation}
\Delta = L_s-\beta L = \frac{L_s}{2\gamma^2} + \frac{R\theta^3}{24}\frac{(R\theta+4y)}{L_s}.
\end{equation}
Since we are interested in cases with large $\gamma$, $\beta$ is taken to be unity except for $(1-\beta) \approx 1/2\gamma^2$ in the leading term. 

Applying Eq.~(7)(8)(9) gives:
\begin{gather}
\mathcal{N}_v = \frac{L_s}{2\gamma^2}+\frac{R\theta^3(3R\theta+4y)}{8L_s}, \\
\mathcal{D} =\frac{L_s}{2\gamma^2}+ \frac{R^2\theta^4}{8L_s}.
\end{gather}

\begin{multline}
w = ke^2 \gamma^2 L_s^2\left(\frac{64[4 L_s^2+\gamma ^2 R\theta^3 (3 R\theta+4 y)]}{(4 L_s^2+\gamma ^2 R^2\theta^4)^3} \right. \\
\left. - \frac{576}{(12L_s^2+\gamma ^2 R\theta^3 (R\theta+4y))^2}\right).   
\label{eq:wyA}
\end{multline}

Eq.~\eqref{eq:wyA} agrees with Eq.~(30) from \cite{Saldin}. The first term, with the cubic in the denominator, comes from the velocity field. The second term comes from the space charge field, and it monotonically decreases with $y$. Note that if $\theta = 0$, the two terms cancel each other, and $w$ vanishes regardless of $\gamma$. There is thus no CSR field for particles on a straight line because the space charge term has been subtracted in Eq.~(6).

\begin{figure}[!h]
	\centering
	\includegraphics[width=0.4\textwidth]{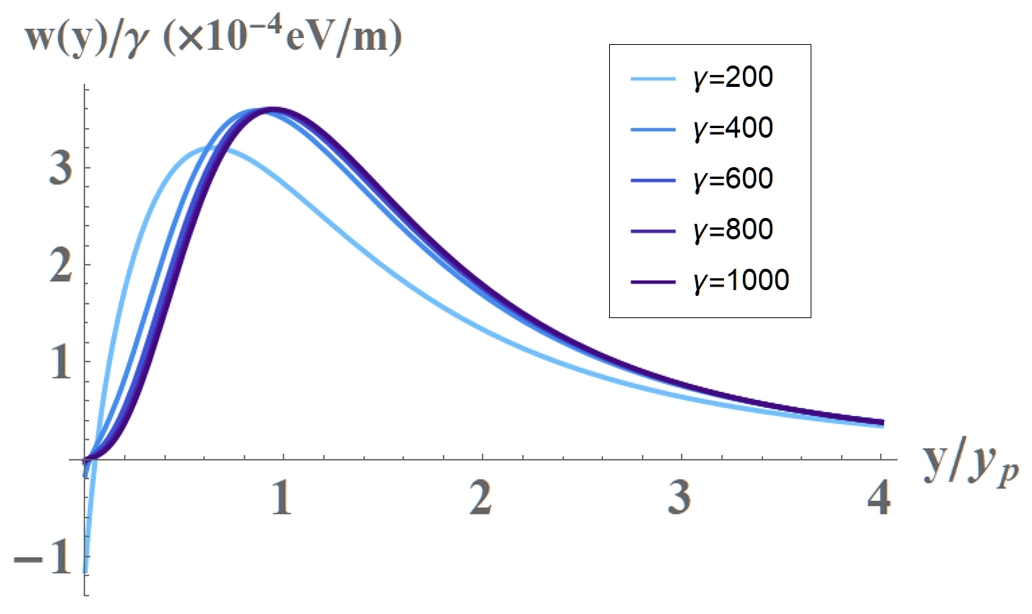}
	\caption{(Color) $w(y)$ for various $\gamma$ values with $R = 1.0$~m and $\theta = 0.04$~rad. For large $\gamma$, $w(y)$ is maximized at $y \approx y_p = \frac{1}{2}\gamma R\theta^2$. Note that $w$ is negative at small $y$ due to the space charge field term.}
	\label{wyAgamma}
\end{figure}

Fig.~\ref{wyAgamma} shows the wake as a function of $y$ for different $\gamma$. At large $\gamma$ the wake is localized in the vicinity $y \sim y_p = \frac{1}{2}\gamma R\theta^2$. This means for large $\gamma$, the main contribution to $W(s)$ comes from electrons with a retarded position $y \gg R\theta$. With this the wake reduces to: 
\begin{equation}
w(y) = 4 ke^2\gamma^2 \left[\frac{64 y^3(y+ \gamma^2 R\theta^3)}{(4y^2+\gamma^2 R^2\theta^4)^3} - \frac{9}{(3y+\gamma^2R\theta^3)^2}\right]. 
\end{equation}

Numerical simulation has shown that this is a good approximation

In the limit of $\gamma^2 \gg L_s/R\theta^3$, the second term in the velocity field term dominates, giving:  
\begin{equation}
w(y) = 256 ke^2\gamma^4 \frac{y^3R\theta^3}{(4y^2+\gamma^2R^2 \theta^4)^3}.
\end{equation}

One can check that $y_p$ maximizes this term. For large $\gamma$ the wake can be approximated as a dirac delta function: 
$w(y) \sim \delta (y-y_p)$. This is a good approximation because the beam distribution over the retarded positions of $P'$ are stretched over a much longer distance than the width of the peak in Fig.~2. To find the corresponding dirac delta function in the $\Delta$ space, we need to calculate $\Delta(y=y_p)$. In the limit $y \gg R\theta$, we have:
\begin{equation}
    \Delta(y) = \frac{y}{2\gamma^2}+\frac{1}{6}R\theta^3-\frac{1}{8y}R^2\theta^4  + O(1/y^2).
\end{equation}

Neglecting higher order terms, we have $\Delta(y_p) = \frac{1}{6}R\theta^3$. Therefore the wake in the $\Delta$ space can be written as:
\begin{equation}
w(\Delta) = A \delta \left(\Delta-\frac{1}{6}R\theta^3\right).
\end{equation}

This makes sense since for large $\gamma$, the dominant term in $\Delta(y)$ is the constant term $R\theta^3/6$. Assuming that the drift in front of the magnet is infinitely long, the normalization factor $A$ can be found by: 
\begin{equation}
A = \int w d\Delta = \int_0^\infty w(y) \frac{\partial \Delta}{\partial y} dy,
\label{eq:AA}
\end{equation}
in which $w(y)$ is given by Eq.(20) and the derivative $\frac{\partial \Delta}{\partial y}$ can be computed from Eq.(21):
\begin{equation}
\frac{\partial \Delta}{\partial y} = \frac{1}{2\gamma^2} + \frac{1}{8y^2}R^2\theta^4.
\end{equation}
\texttt{Mathematica} gives $A = 4ke^2/R\theta $. For a bunch with a longitudinal distribution $\lambda(s)$, the wake expression $W(s)$ is: 
\begin{align}
W(s) &= \int w(s-s')\lambda(s') ds'  \label{eq:general}\\
&= \frac{4ke^2}{R\theta} \lambda\left(s-\frac{1}{6}R\theta^3\right). \label{eq:wyA2} 
\end{align}

Eq.~\eqref{eq:wyA2} agrees with Eq.~(10) in \cite{Emma}. Assuming the bunch has a bunch length of $l_b$, then for a large entrance angle $\theta$ such that $R\theta^3/6 \gg l_b$, the wake attenuates. In the case where the bunch has not entirely entered the magnet (i.e. $R\theta < l_b$), then the tail portion of the bunch outside the magnet sees no CSR wake.

There are some limitations in applying the wake expression in Eq.~\eqref{eq:wyA2}. First, the derivation assumes that $\gamma^2 \gg L_s/R\theta^3$, which implies that $\theta \gg 1/\gamma$ is required. Therefore depending on the $\gamma$ value, the observation point cannot be too close to the magnet entrance. For small $\theta$ such that $\theta \sim 1/\gamma$, or equivalently $\Delta \sim R/\gamma^3$, the wake needs to be found numerically. However for a very large $\gamma$ such a small numerical scale can be difficult to resolve in simulation software. A numerical formulation to resolve this has been presented in IPAC 2017 \cite{Chris_IPAC}. 

In addition, the derivation assumes a drift with a length $y_m \gg y_p$, so that most of the peak in Fig.~(2) appears within the drift. Note that $y_p$ can be large since it increases with $\gamma$. For an insufficiently long drift, one needs to use the more general formula from Eq.~\eqref{eq:general} to calculate $W(s)$:
\begin{eqnarray}
\begin{aligned}
W(s) &= \int^{s-\Delta(y=0)}_{s-\Delta(y=y_m)} w(s-s')\lambda(s') ds' \\
&= \int^{y_m}_{0} w(y)\lambda(s-\Delta(y))\frac{\partial \Delta}{\partial y} dy,
\label{eq:wym}
\end{aligned}
\end{eqnarray}
in which $w(y)$ is given by Eq.\eqref{eq:wyA}. Note that change of variable from $s'$ to $y$ has been applied so there is no need to find an expression for $w(\Delta)$. In the case with $y_m \ll y_p$, contribution from the drift becomes insignificant for $\gamma \theta \gg 1$. 

\subsubsection{Case C}
The geometry for case $C$ is shown in Fig.~\ref{caseC} below. Since case C is an extension of case A, we will apply a similar derivation process. 
\begin{figure}[!h]
	\centering
	\includegraphics[width=0.25\textwidth]{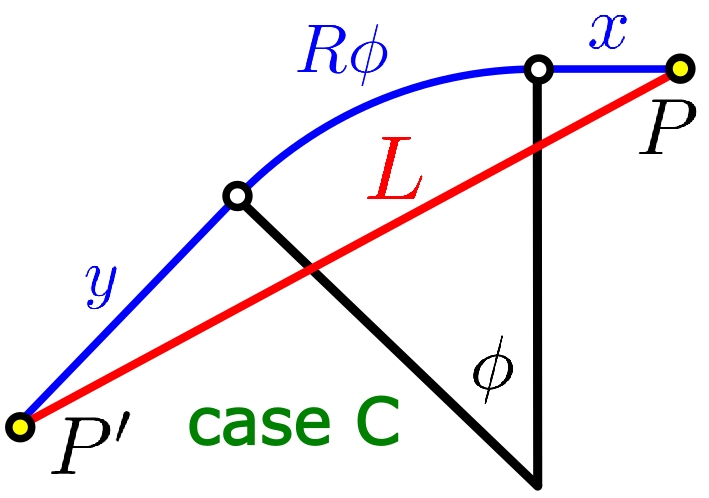}
	\caption{(Color) Geometry for case C.}
	\label{caseC}
\end{figure}

To find $\mathcal{N}_v$ and $\mathcal{D}$ we have to first fix our coordinate system. Let us pick the exit point of the bending magnet to be the origin, and the exit drift to lie along the $+x$ axis. This choice will be shown convenient for case E too. Let $\phi$ denote the $total$ bending angle of the magnet. Vector analysis shows that:
\begin{gather}
\textbf{n} = \langle 1,0,0 \rangle,\\
\textbf{n}' = \langle \cos{\phi},\sin{\phi},0 \rangle,\\
L_x=  y\cos{\phi} + L_c\cos{(\phi/2)} + x,\\
L_y = y\sin{\phi} + L_c\sin{(\phi/2)}.
\end{gather}

We again expand all quantities in small $\phi$, holding $R\phi$ constant, up to order of $\phi^2$. This yields:
\begin{gather}
L_c = R\phi - R\phi^3/24, \\
L = L_s - \frac{\phi^2}{24}\frac{R^2\phi^2+4 R \phi(x+y)+12xy}{L_s},\\
\Delta = \frac{L_s}{2\gamma^2} + \frac{\phi^2}{24}\frac{R^2\phi^2+4 R \phi(x+y)+12xy}{L_s},
\end{gather}
in which $L_s = R\phi+x+y$ for case C. Applying Eq.~(7) and Eq.~(8) gives:
\begin{gather}
\mathcal{N}_v = \frac{L_s}{2\gamma^2}+\frac{(R\phi+2x)(3R\phi+2x+4y)\phi^2}{8L_s}, \\
\mathcal{D} =\frac{L_s}{2\gamma^2}+ \frac{(R\phi + 2x)^2\phi^2}{8L_s}.
\end{gather}

These lead to:
\begin{multline}
w = ke^2 \gamma^2  L_s^2\left(\frac{64[4 L_s^2+\gamma ^2 \phi^2 (R\phi +2 x) (3 R\phi+2x+4 y)]}{(4 L_s^2+\gamma ^2 \phi^2 (R\phi+2 x)^2)^3} \right.\\
\left.- \frac{576}{(12L_s^2+\gamma ^2 \phi^2 (R^2\phi ^2+4R\phi (x+y)+12xy))^2}\right).   
\label{eq:wyC}
\end{multline}

Eq.~\eqref{eq:wyC} agrees with Eq.~(34) from \cite{Saldin}. As expected, the wake vanishes if $\phi \rightarrow 0$ or $x \rightarrow \infty$. In the limit $\gamma^2 \gg L_s/(R\phi+2x)\phi^2$, the second term in the velocity field term dominates. With the additional limit of $y \gg R\phi$ and $y \gg x$, the wake reduces to:
\begin{equation}
w = 256 ke^2\gamma^4 \frac{y^3 (R\phi+2x)\phi^2}{(4y^2+\gamma^2(R\phi+2x)^2 \phi^2)^3}.
\label{eq:wyC2}
\end{equation}

This wake is maximized at $y_p = \frac{1}{2}\gamma (R\phi+2x)\phi$. In the limit of large $y$ we also have:
\begin{equation}
    \Delta(y) = \frac{y}{2\gamma^2}+\frac{1}{6}(R\phi+3x)\phi^2-\frac{1}{8y}(R\phi+2x)^2 \phi^2 + O(1/y^2).
\end{equation}
Similar to case A, we assume that $y_m$, the length of the drift in front of the magnet, is much longer than $y_p$. With $\Delta(y_p) = \frac{1}{6}(R\phi+3x)\phi^2$, the wake can be approximated as a dirac delta function:
\begin{equation}
w(\Delta) = A \delta \left(\Delta-\frac{1}{6}(R\phi+3x)\phi^2\right).
\end{equation}

The normalization factor $A$ can be found by: 
\begin{equation}
A = \int_0^\infty w(y) \frac{\partial \Delta}{\partial y} dy = \frac{4ke^2}{R\phi+2x},
\end{equation}

For a bunch with a longitudinal distribution $\lambda(s)$, the wake expression $W(s)$ is: 
\begin{eqnarray}
\begin{aligned}
W(s) &= \int w(s-s')\lambda(s') ds' \\
&= \frac{4ke^2}{(R\phi+2x)} \lambda\left(s-\frac{1}{6}(R\phi+3x)\phi^2\right),
\end{aligned}
\label{eq:wyC3}
\end{eqnarray}

which agrees with Eq.~(10) in \cite{Emma}. For $x \rightarrow 0$, we recover the expression for case A. For large $x$, the wake vanishes as expected. Since the derivation assumes $\gamma^2 \gg L_s/(R\phi+2x)\phi^2$, the bending angle $\phi$ cannot be too small. Also, the drift in front of the magnet has to be much longer than $y_p$. Otherwise one needs to use a more general formula as Eq.~\eqref{eq:wym}.

\subsubsection{Case E1}
\begin{figure}[!h]
	\centering
	\includegraphics[width=0.4\textwidth]{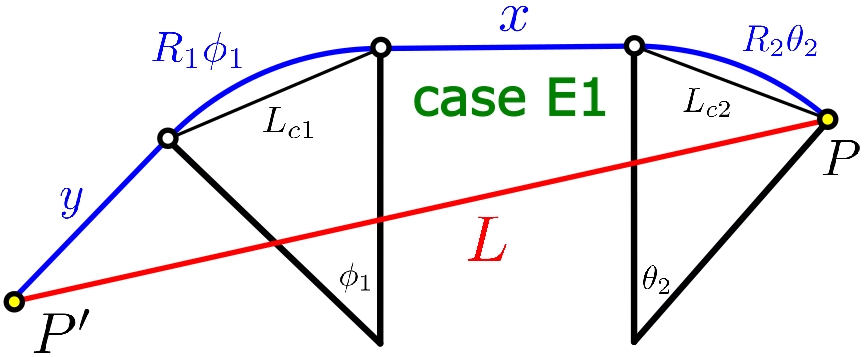}
	\caption{(Color) Geometry for case E1.}
	\label{caseE1}
\end{figure}

Case E has two subcases depending on the bending direction of the second magnet. For case E1, the two magnets have the same bending direction. For opposite bending directions, see case E2 in the following subsection. The geometry for case E1 is shown in Fig.~\ref{caseE1}. A second magnet with a bending radius $R_2$ has been introduced at a distance $x$ behind the first magnet. The observation point $P$ is located at an angle $\theta_2$ into the second magnet. Note that $\theta_2$ is $not$ the total bending angle of the second magnet.

Since case E is an extension of case C, we will apply a similar derivation procedure. Using the same coordinate system as in case C, we now have:
\begin{gather}
\textbf{n} = \langle \cos{\theta_2},-\sin{\theta_2},0 \rangle,\\
\textbf{n}' = \langle \cos{\phi_1},\sin{\phi_1},0 \rangle,\\
L_x =  (y\cos{\phi_1} + L_{c1}\cos{\frac{\phi_1}{2}} +x+L_{c2}\cos{\frac{\theta_2}{2}}), \\
L_y = (y\sin{\phi_1} + L_{c1}\sin{\frac{\phi_1}{2}}-L_{c2}\sin{\frac{\theta_2}{2}}),
\end{gather}

in which $L_{c1}$ and $L_{c2}$ are respectively the chord length associated with $R_1 \phi_1$ and $R_2 \theta_2$. Expanding in small $\phi_1$ and $\theta_2$ gives:
\begin{multline}
L = L_s - \frac{1}{24 L_s}\Big\{[R_1^2\phi_1^2+12(R_2\theta_2+x)y \\
+ 4R_1\phi_1 (R_2\theta_2+x+y)\phi_1^2] + 6[R_2\theta_2(R_1\phi_1+2y)\phi_1\theta_2] \\
+ [R_2\theta_2(4R_1\phi_1+R_2\theta_2+4x+4y)]\theta_2^2  \Big\},
\end{multline}

in which $L_s = (R_1\phi_1+R_2\theta_2+x+y$) for case E1. It follows that:
\begin{multline}
\mathcal{N}_v = \frac{L_s}{2\gamma^2}+\frac{1}{8L_s}
[(3R_1\phi_1+2x+4y+2R_2\theta_2)\phi_1 \\
+ (4R_1\phi_1+4x+4y+3R_2\theta_2)\theta_2] \\
\times (R_1\phi_1^2+2x\phi_1+2R_2\phi_1\theta_2+R_2\theta_2^2),
\end{multline}
\begin{equation}
\mathcal{D}=\frac{L_s}{2\gamma^2}+\frac{1}{8L_s}(R_1\phi_1^2+2x\phi_1+2R_2\phi_1\theta_2+R_2\theta_2^2)^2.
\end{equation}

In the limit of
\begin{equation}
\gamma^2 \gg L_s/(\phi_1+\theta_2)(R_1\phi_1^2+2x\phi_1+2R_2\phi_1\theta_2+R_2\theta_2^2), 
\label{gammaE1}
\end{equation}
the dominant term again comes from the velocity field. As in case C, we take the additional limit of large $y$ (i.e. $y \gg R_1\phi_1,x, R_2\theta_2$), and the wake reduces to:
\begin{equation}
w = \frac{256ke^2\gamma^4 y^3(\phi_1+\theta_2)(R_1\phi_1^2+2x\phi_1+2R_2\phi_1\theta_2+R_2\theta_2^2)}{[4y^2+\gamma^2(R_1\phi_1^2+2x\phi_1+2R_2\phi_1\theta_2+R_2\theta_2^2)^2]^3}.
\label{eq:wyE}
\end{equation}

One can verify that $y_p = \frac{1}{2}\gamma (R_1\phi_1^2+2x\phi_1+2R_2\phi_1\theta_2+R_2\theta_2^2)$ maximizes $w(y)$ in Eq.~\eqref{eq:wyE}. If the second magnet is removed (i.e. $\theta_2 = 0$), Eq.~\eqref{eq:wyE} reduces to Eq.~\eqref{eq:wyC2} in case C. In the limit of large $y$ we also have:
\begin{multline}
\Delta = \frac{y}{2\gamma^2} + \frac{1}{6}\left[(R_1\phi_1+3x)\phi_1^2+R_2\theta_2(3\phi_1^2+3\phi_1\theta_2+\theta_2^2)\right] \\-\frac{1}{8y}(R_1\phi_1^2+2x\phi_1+2R_2\phi_1\theta_2+R_2\theta_2^2)^2+ O(1/y^2).
\end{multline}
As before, the constant term in $\Delta$ gives the location of the dirac delta function. The normalization factor $A$ is:
\begin{equation}
A = \int_0^\infty w(y) \frac{\partial \Delta}{\partial y} dy 
= \frac{4ke^2(\phi_1+\theta_2)}{R_1\phi_1^2+2x\phi_1+2R_2\phi_1\theta_2+R_2\theta_2^2}.
\end{equation}

So the approximated wake expression for case E1 is:
\begin{multline}
W(s) = \frac{4ke^2(\phi_1+\theta_2)}{R_1\phi_1^2+2x\phi_1+2R_2\phi_1\theta_2+R_2\theta_2^2} \times\\ \lambda ( s - \frac{1}{6}\left[(R_1\phi_1+3x)\phi_1^2+R_2\theta_2(3\phi_1^2+3\phi_1\theta_2+\theta_2^2)\right]).
\label{eq:wyC4}
\end{multline}

Let us examine some limiting cases. First, if we set $\theta_2 =0$, we recover the $W(s)$ for case C as expected (See Eq.~\eqref{eq:wyC3}). If we instead remove the first magnet by setting $\phi_1 = 0$, we recover case A with only the second magnet (See Eq.~\eqref{eq:wyA2} with $\theta \rightarrow \theta_2$). Another interesting limit is to set $x=0$ and $R_1=R_2=R$. This physically corresponds to removing the drift between the two magnets and merging the two magnets. The result is:
\begin{equation}
W(s) = \frac{4ke^2}{R(\phi_1+\theta_2)}  \lambda\left(s-\frac{1}{6}R(\phi_1+\theta_2)^3\right).
\end{equation}
As expected, we recover the $W(s)$ for case A with the observation point located at $\theta=(\phi_1+\theta_2)$ into the merged magnet. Similar to case C and case A, there are limitations in applying Eq.~\eqref{eq:wyC4}. The length of the drift in front of the first magnet has to be much greater than $y_p$. Also, the assumption of large $\gamma$ in Eq.~\eqref{gammaE1} requires $(\phi_1 + \theta_2)$ to be not small.

\subsubsection{Case E2}
In contrast to case E1, the second magnet has a bending direction $opposite$ to the first magnet. With our coordinate system the vector $\textbf{n}$ now becomes $\langle \cos{\theta_2},\sin{\theta_2},0 \rangle$. Following the same derivation procedure as in case E1, we obtain: 
\begin{multline}
W(s) = \frac{4ke^2(\phi_1-\theta_2)}{R_1\phi_1^2+2x\phi_1+2R_2\phi_1\theta_2-R_2\theta_2^2} \times\\ 
\lambda \left( s -\frac{1}{6}\left[(R_1\phi_1+3x)\phi_1^2+R_2\theta_2(3\phi_1^2-3\phi_1\theta_2+\theta_2^2)\right] \right).   
\end{multline}

As expected, the result is very similar to case E1, with $\theta_2 \rightarrow -\theta_2$. However, the term $R_2 \theta_2$ remains unchanged since it comes from the path length and is always positive. 

\subsubsection{Case G1 and G2}
\begin{figure}[!h]
	\centering
	\includegraphics[width=0.4\textwidth]{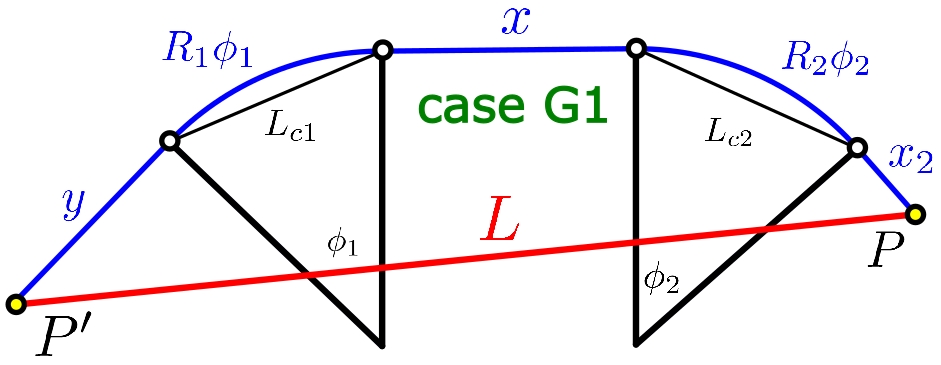}
	\caption{(Color) Geometry for case G1.}
	\label{caseG1}
\end{figure}

Similar to case E, there are two subcases in case G depending on the two bending directions. The geometry for case G1 is shown in Fig.~\ref{caseG1}. The observation point $P$ is now located a distance $x_2$ down the second magnet. Let $\phi_2$ denote the total bending angle of the second magnet. 

With the same derivation procedure as in case E1 and E2, we obtain: 

\begin{multline}
W(s) = \frac{4ke^2(\phi_1 \pm \phi_2)}{R_1\phi_1^2+2x\phi_1+2R_2\phi_1\phi_2 \pm R_2\phi_2^2 +2x_2(\phi_1 \pm \phi_2)} \\
\times \lambda (s -\frac{1}{6}[(R_1\phi_1+3x)\phi_1^2+R_2\phi_2(3\phi_1^2 \pm 3\phi_1\phi_2+\phi_2^2)  \\ 
+ 3x_2(\phi_1 \pm \phi_2)^2]).
\label{eq:WG}
\end{multline}

Note that the $\pm$ sign denotes $``+"$ for case G1 (when the two magnets bend in the same direction) and $``-"$ for case G2 (with opposite bending directions). One can verify that for $x_2 = 0$, we recover the expressions for case E1 and E2. If we merge the two magnets by taking $x=0$ and $R_1=R_2 = R$, then case G1 reduces to case C with $\phi = \phi_1+ \phi_2$, as expected.

\subsection{Even cases}
We will first re-derive the wake expression for case B and D, then apply a similar formulation to obtain the expressions for case F and H. 

\subsubsection{Case B}
\begin{figure}[!h]
	\centering
	\includegraphics[width=0.20\textwidth]{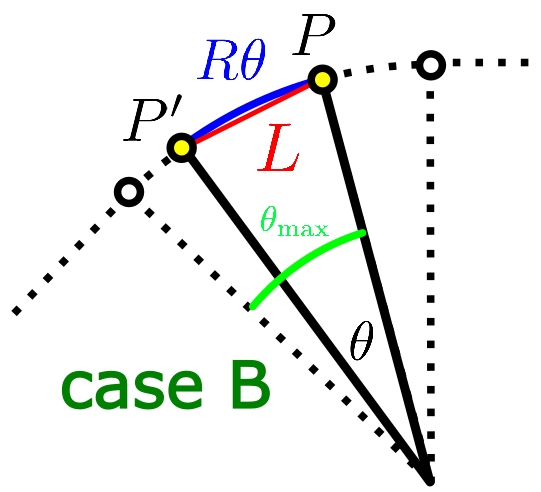}
	\caption{(Color) Geometry for case B. Note that the $\theta$ here is $not$ the same as the $\theta$ defined in case A. The $\theta_{\text{max}}$ here corresponds to the $\theta$ in case A.}
	\label{caseB}
\end{figure}

The geometry for case B is shown in Fig.~\ref{caseB}. The point $P'$ and $P$ are inside the same magnet separated by an angle $\theta$. 

Since $P'$ is inside a bend, the tail electron at the retarded time has an centripetal acceleration with magnitude $|\textbf{a}'|=\beta^2 c^2/R$. This adds the first extra term to the wake:
\begin{equation}
\begin{aligned}
w &= ke^2 \left(\frac{\mathcal{N}_a}{\mathcal{D}^3}+ \frac{1}{\gamma^2} \left(\frac{\mathcal{N}_v}{\mathcal{D}^3} - \frac{1}{\Delta^2}\right)\right),
\end{aligned}
\end{equation}

in which $\mathcal{N}_a \equiv  \beta\textbf{n} \boldsymbol{\cdot} (\textbf{L} \times [(\textbf{L}-L\beta\textbf{n}') \times \textbf{a}'])/c^2$. We first fix our coordinate system by choosing $\textbf{n} = \langle 1,0,0 \rangle$, then we have:
\begin{gather}
\textbf{n}' = \langle \cos{\theta}, \sin{\theta}, 0 \rangle, \\
\textbf{a}' = \frac{\beta^2 c^2}{R}\langle -\sin{\theta},\cos{\theta},0 \rangle,\\
\textbf{L} = \langle (L_c\cos{(\theta/2)}), (L_c\sin{(\theta/2)}),0\rangle.
\end{gather}
Expanding in small $\theta$, keeping $R\theta$ constant, gives: 
\begin{gather}
L = L_s - \frac{R \theta^3}{24}, \\
\mathcal{N}_a = \frac{-L_s}{2\gamma^2}\frac{(R\theta)}{2R}\theta+\frac{(R\theta)^2}{16R}\theta^3,\\
\mathcal{N}_v = \frac{L_s}{2\gamma^2}+\frac{3R\theta^3}{8},\\
\mathcal{D} = \frac{L_s}{2\gamma^2}+\frac{R\theta^3}{8},
\end{gather}
in which $L_s = R\theta$ for case B. Note that while $even$ terms in $\theta$ survive in $L$, $\mathcal{N}_v$, and $\mathcal{D}$, $odd$ terms survive in $\mathcal{N}_a$, and it is required to keep the expansion up to order $\theta^3$ in $\mathcal{N}_a$. The expression for $\mathcal{N}_a$ is $not$ simplified to explicitly show this observation, and the $R$ in the denominator comes directly from $\textbf{a}'$ and therefore does not have a $\theta$ associated with it. It follows that:

\begin{multline}
w = \frac{32ke^2\gamma^4}{R^2}\left[ \frac{\gamma^2 \theta^2-4}{(\gamma^2 \theta^2+4)^3}\right. \\ 
\left. +\frac{2}{\gamma^2\theta^2}\left(\frac{3\gamma^2 \theta^2+4}{(\gamma^2 \theta^2+4)^3} - \frac{9}{(\gamma^2 \theta^2 + 12)^2} \right) \right],
\end{multline}

which agrees with Eq.(36) in \cite{Saldin}. For $\gamma \theta \ll 1$, the wake reduces to:
\begin{equation}
w(\gamma \theta \ll 1) = \frac{-4ke^2\gamma^4}{3R^2}.
\end{equation}
The physical significance of this result has been discussed in \cite{Saldin}. Here we are interested in the case with $\gamma \theta \gg 1$, for which the acceleration field term dominates:
\begin{equation}
w (\gamma \theta \gg 1) = \frac{32ke^2}{R^2\theta^4}.
\end{equation}

In the limit of large $\gamma$ we also have:
\begin{equation}
\Delta = L_s - \beta L = \frac{L_s}{2\gamma^2}+ \frac{R \theta^3}{24} \rightarrow \frac{R \theta^3}{24}
\end{equation}
To solve for $W(s)= \int w(s-s')\lambda(s')ds'$ one can simply write down $w(\Delta)$ by inverting Eq.(65). However, for later cases with more complicated geometry, $w(\Delta)$ cannot be easily inverted. As shown in \cite{Emma}, a more general method is to apply integration by parts. This requires us to find the function $u$ such that $\partial u/\partial s' = w(s-s')$. Since both $w$ and $\Delta$ are functions of the independent variable $\theta$, we can compute:
\begin{equation}
\frac{du}{d\theta} = \frac{\partial u}{\partial s'}\frac{ds'}{d\theta} = -w \frac{d\Delta}{d\theta} = \frac{-4ke^2}{R\theta^2},  
\end{equation}
in which the minus sign is present because $\Delta = s-s'$. Integration over $\theta$ gives $u = 4ke^2/R\theta$, which turns out to be the same as the normalization factor $A$ in case A (see Eq.~\eqref{eq:AA}). Note that the limit of integration is $(s-\Delta_{\text{max}}) < s' < s $, in which $\Delta_{\text{max}} = \Delta(\theta=\theta_{\text{max}}) = R\theta_{\text{max}}^3/24$. When $\theta=\theta_{\text{max}}$, the point $P'$ is located at the entrance of the magnet. One can also interpret $\theta_{\text{max}}$ as the entrance angle of the observation point $P$, which is exactly the ``$\theta$'' defined in case A. For $s'<(s-\Delta_{\text{max}})$, one needs to use results from part A. It follows that:
\begin{equation}
\begin{aligned}
W(s) &= (u\lambda)\bigg |^s_{s-\Delta_{\text{max}}} -\int^s_{s-\Delta_{\text{max}}} u\frac{\partial \lambda(s')}{\partial s'} ds'\\
& = -4ke^2 \left[ \frac{\lambda(s-\Delta_{\text{max}})}{R\theta_{\text{max}}} + \int^s_{s-\Delta_{\text{max}}} \frac{1}{R\theta} \frac{\partial \lambda(s')}{\partial s'} ds'\right],
\label{eq:wB}
\end{aligned}
\end{equation}

which agrees with Eq.~(6) in \cite{Emma}. To apply this expression, it is required that $\gamma \theta \gg 1$, so the observation point $P$ cannot be too close to the magnet's entrance. The expression overlooks the wake contributions from $\theta \sim 1/\gamma$ or even smaller $\theta$. As discussed in case A, these contributions can be calculated numerically. For a very large $\gamma$ these contributions can be ignored.

\subsubsection{Case D}
\begin{figure}[!h]
	\centering
	\includegraphics[width=0.25\textwidth]{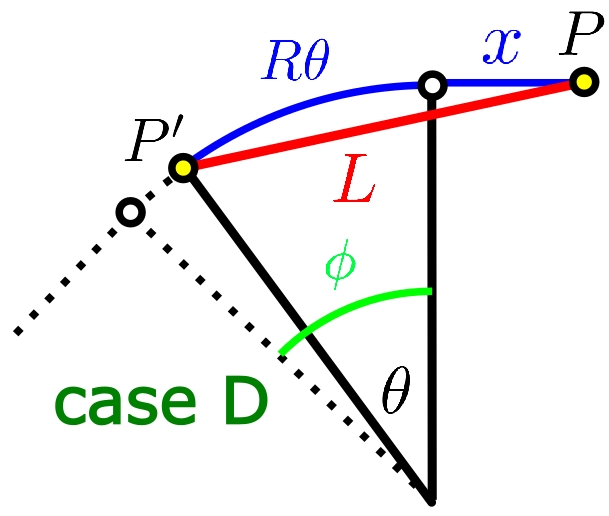}
	\caption{(Color) Geometry for case D. Note that the $\theta$ here is $not$ the $\theta$ defined in either case A or B.}
	\label{caseD}
\end{figure}

The geometry for case D is shown in Fig.~\ref{caseD}. The point $P'$ is located at an angle $\theta$ measured from the $exit$ of the magnet. 

Since Case D is an extension of case B, we will apply a similar derivation process. Let us again choose $\textbf{n} = \langle 1,0,0\rangle$, then $\textbf{n}'$ and $\textbf{a}'$ are the same as in case B, and:
\begin{gather}
\textbf{L} = \langle (L_c\cos{(\theta/2)} + x), (L_c\sin{(\theta/2)}),0\rangle.
\end{gather}
Expanding in small $\theta$ gives: 
\begin{gather}
L = L_s - \frac{R \theta^3}{24}\frac{ (R \theta + 4x)}{L_s}, \\
\mathcal{N}_a = \frac{-L_s}{2\gamma^2}\frac{\theta^2}{2}+\frac{\theta^4}{16}
\frac{(R\theta+2x)^2}{(R\theta+x)},\\
\mathcal{N}_v = \frac{L_s}{2\gamma^2}+\frac{(3R^2\theta^2 + 8R\theta x +4x^2)\theta^2}{8(R\theta+x)},\\
\mathcal{D} = \frac{L_s}{2\gamma^2}+\frac{(R\theta+2x)^2\theta^2}{8L_s},
\end{gather}
in which $L_s = R\theta +x$ in case D. One can verify that as $x \rightarrow 0$, these quantities reduce to the ones in case B. 
It follows that:
\begin{multline}
w = 32ke^2\gamma^4 L_s^2
\left[
\frac{(\gamma^2 \theta^2(R\theta+2x)^2-4L_s^2)\theta^2}{[\gamma^2 \theta^2(R\theta+2x)^2+4L_s^2]^3} \right.
\\+\frac{2}{\gamma^2} \left( \frac{\gamma^2\theta^2(3R^2\theta^2+8R\theta x +4x^2)+4L_s^2}{[\gamma^2\theta^2(R\theta+2x)^2+4L_s^2]^3} \right.
\\ \left. \left.- \frac{9}{[\gamma^2\theta^2R\theta(R\theta+4x)+12L_s^2]^2}\right) \right],  
\end{multline}

which agrees with Eq.(36) in \cite{Saldin}. If we set $x =0$, the wake reduces to the one in case B. To focus on case D we assume $x>0$ from now on. In the limit $\theta \rightarrow 0$, the wake vanishes:
\begin{equation}
w(\theta \rightarrow 0) = \left[0+\frac{4\gamma^2}{x^2}-\frac{4\gamma^2}{x^2}\right] =0,
\end{equation}
regardless of the value of $\gamma$. As in case B, we are interested in the limit $\gamma \theta \gg 1$, for which the acceleration field term dominates:
\begin{equation}
w(\gamma\theta \gg 1) = 32ke^2 \frac{(R\theta+x)^2}{\theta^2(R\theta +2x)^4}.
\end{equation}
In a more strict limit of $\gamma^2 \gg  (R\theta+x)/R\theta^3$ we have:
\begin{equation}
\Delta = \frac{R \theta^3}{24}\frac{ (R \theta + 4x)}{R\theta +x}.
\end{equation}
To solve for $W(s)$ we need to find the function $u$ such that $\partial u/\partial s' = w(s-s')$. Similar to case B, we have:
\begin{equation}
\frac{du}{d\theta} = \frac{\partial u}{\partial s'}\frac{ds'}{d\theta} = -w \frac{d\Delta}{d\theta} = \frac{-4ke^2R}{(R\theta+2x)^2},  
\end{equation}
Integration over $\theta$ gives $u = 4ke^2/(R\theta+2x)$. We again observe that $u$ is the same quantity as the normalization factor $A$ in case C. We are now ready to apply integration by parts to find $W(s)$. 
Note that the region of integration is $(s-\Delta_{\text{max}}) < s' < (s-\Delta_{\text{min}})$, in which $\Delta_{\text{max}} = \Delta(\theta=\theta_{\text{max}})$ and $\Delta_{\text{min}} = \Delta(\theta=0) = 0$. Since $\theta$ in case D is measured from the $exit$ of the magnet, we have $\theta_{\text{max}} = \phi$, the total bending angle of the magnet. It follows that:
\begin{equation}
\begin{aligned}
W(s) &= (u\lambda)\bigg |^s_{s-\Delta_{\text{max}}} -\int^s_{s-\Delta_{\text{max}}} u\frac{\partial \lambda(s')}{\partial s'} ds'\\
& = -4ke^2 \left[ \frac{\lambda(s-\Delta_{\text{max}})}{R\phi+2x} + \int^s_{s-\Delta_{\text{max}}} \frac{1}{R\theta+2x} \frac{\partial \lambda(s')}{\partial s'} ds'\right].
\end{aligned}
\label{eq:wyD1}
\end{equation}

The result agrees with Eq.~(15) \cite{Emma}. Note that one of the boundary term $u\lambda$ ($s' \rightarrow s$) vanishes since $w(\theta \rightarrow 0) = 0$. One can check that if the observation point $P$ is located at the exit of the magnet (i.e. $x=0, \phi = \theta_{\text{max}}$), we recover the result of case B. For large $x$, the expression vanishes as expected. When using equation Eq.~\eqref{eq:wyD1}, one should abide to the limit $\gamma^2 \gg  (R\theta+x)/R\theta^3$. This means for a very large $x$ value, $W(s)$ can give inaccurate results.

Otherwise 
At eq 27, mention ignorance of small angle

\subsubsection{Case F1}
\begin{figure}[!h]
	\centering
	\includegraphics[width=0.4\textwidth]{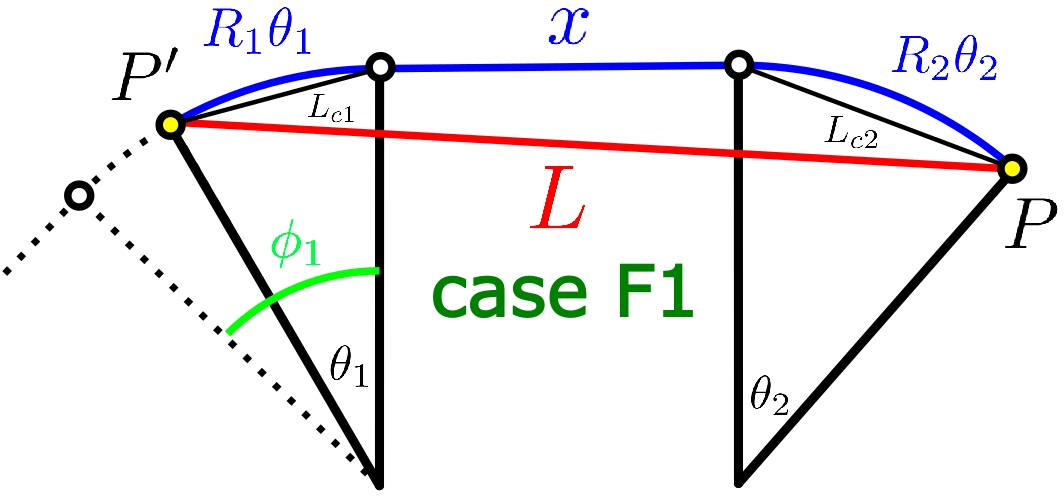}
	\caption{(Color) Geometry for case F1.}
	\label{caseF1}
\end{figure}

Fig.~\ref{caseF1} shows the geometry for case F1. In terms of the location of the observation point, case F1 and E1 are the same. Similar to case E1, case F1 has two magnets bending in the same direction. For opposite bending directions, see case F2 in the following subsection. The geometry for case F1 is shown in Fig.~\ref{caseF1}.

Since case F is an extension of case D, we follow similar derivations. With the coordinate system defined in case E1, we now have:
\begin{gather}
\textbf{n} = \langle \cos{\theta_2},-\sin{\theta_2},0 \rangle,\\
\textbf{n}' = \langle \cos{\theta_1},\sin{\theta_1},0 \rangle,\\
\textbf{a}' = \frac{\beta^2 c^2}{R}\langle -\sin{\theta},\cos{\theta},0 \rangle,\\
L_x = (L_{c1}\cos{\frac{\theta_1}{2}} +x+L_{c2}\cos{\frac{\theta_2}{2}})\\
L_y = (L_{c1}\sin{\frac{\theta_1}{2}}-L_{c2}\sin{\frac{\theta_2}{2}}).
\end{gather}
From case B and D we know for large $\gamma$ the dominant term in $w$ is the acceleration field term:
\begin{multline}
w =\frac{32ke^2\gamma^4 L_s^2}{R_1}(R_1\theta_1^2+2R_1 \theta_1\theta_2 +2x\theta_2+R_2\theta_2^2) \times \\
\frac{\gamma^2 (R_1\theta_1^2+2R_2\theta_1\theta_2+2x\theta_1+R_2\theta_2^2)^2 - 4L_s^2}{[\gamma^2 (R_1\theta_1^2+2R_2\theta_1\theta_2+2x\theta_1+R_2\theta_2^2)^2 +4 L_s^2 ]^3}, 
\end{multline}

in which $L_s = (R_1\theta_1 + x + R_2\theta_2)$ for case F. Note that $w (x \rightarrow \infty) \propto 1/x \rightarrow 0$ as expected. Assuming $R_1$ and $R_2$ are on the same order of magnitude ($R_1 \approx R_2 \approx R$), then in the limits of $\gamma (\theta_1+\theta_2)  \gg 1$ we have:
\begin{equation}
w =\frac{32ke^2L_s^2}{R_1}
\frac{(R_1\theta_1^2+2R_1 \theta_1\theta_2 +2x\theta_2+R_2\theta_2^2)}{(R_1\theta_1^2+2R_2\theta_1\theta_2+2x\theta_1+R_2\theta_2^2)^4},
\end{equation}

Note that the numerator and denominator of $w$ do $not$ cancel. With a more strict limit of $\gamma^2 \gg L_s/R(\theta_1+\theta_2)^3$ we have:
\begin{multline}
\Delta = \frac{1}{24L_s}[R_1\theta_1^3(R_1\theta_1+4x+4R_2\theta_2)\\+6R_1\theta_1^2 R_2 \theta_2^2 +R_2\theta_2^3(4R_1\theta_1+4x+R_2\theta_2)], 
\end{multline}

To find $W(s)$ we again look for the function $u$ such that $\partial u/\partial s' = w(s-s')$. It follows that:
\begin{gather}
\frac{du}{d\theta_1} = -4ke^2\frac{(R_1\theta_1^2+2R_1 \theta_1\theta_2 +2x\theta_2+R_2\theta_2^2)}{(R_1\theta_1^2+2R_2\theta_1\theta_2+2x\theta_1+R_2\theta_2^2)^2},\\
u=\frac{4ke^2(\theta_1+\theta_2)}{R_1\theta_1^2+2R_2\theta_1\theta_2+2x\theta_1+R_2\theta_2^2}.
\end{gather}
Again, we see that $u$ equals to $A$ in the pairing case E1. Integration by parts gives:
\begin{multline}
W(s) = -4ke^2 \left[
-\frac{\lambda(s-\Delta_{\text{min}})}{R_2\theta_2} \right.\\+
\frac{(\phi_1+\theta_2)\lambda(s-\Delta_{\text{max}})}{R_1\phi_1^2+2R_2\phi_1\theta_2+2x\phi_1+R_2\theta_2^2}\\ 
\left. +\int^{s-\Delta_{\text{min}}}_{s-\Delta_{\text{max}}} \frac{(\theta_1+\theta_2)}{R_1\theta_1^2+2R_2\theta_1\theta_2+2x\theta_1+R_2\theta_2^2} \frac{\partial \lambda(s')}{\partial s'} ds' \right], 
\label{eq:wF1}
\end{multline}

in which $\Delta_{\text{max}} = \Delta(\theta_1=\phi_1,\theta_2)$ and
\begin{equation}
\Delta_{\text{min}} = \Delta(\theta_1=0,\theta_2) = \frac{R_2\theta_2^3}{24}\frac{R_2\theta_2+4x}{R_2\theta_2+x}.
\end{equation}

The first term in $W(s)$, which corresponds to the integration boundary at $\theta_1 = 0$, might look physically invalid since it diverges for $\theta_2 \rightarrow 0$ and does $not$ vanish for $x \rightarrow \infty$. However due to the required limit of $\gamma^2 \gg L_s/R(\theta_1+\theta_2)^3$, one simply can $not$ apply Eq.~\eqref{eq:wF1} with a very small $\theta_2$ or large $x$. For $\theta_2 \rightarrow 0$, one should simply apply case D. For $x \rightarrow \infty$ we know the wake vanishes, and the two magnets should be considered ``decoupled". 
To test the validity of Eq.~\eqref{eq:wF1}, let us consider a case with $x=0$ and $R_1 = R_2=R$. This equivalently merges the two magnets into one, and the point $P$ is located ($\phi_1+\theta_2$) into the merged magnet. To compute the $W(s)$ at point P due to the merged magnet, we need to add the contribution from the first sub-magnet using case F and the second sub-magnet using case B. Applying Eq.~\eqref{eq:wF1} and  Eq.~\eqref{eq:wB} respectively gives:
\begin{multline}
W(s) = W_{\text{B}}(s,\theta_{\text{max}}=\theta_2) + W_{\text{F1}}(s,x=0)
\\= -4ke^2 \left[ \frac{\lambda(s-\Delta_{\text{B,max}})}{R\theta_2} + \int^s_{s-\Delta_{\text{B,max}}} \frac{1}{R\theta} \frac{\partial \lambda(s')}{\partial s'} ds'\right]\\
- 4ke^2 \left[
-\frac{\lambda(s-\Delta_{\text{F,min}})}{R_2\theta_2} +
\frac{\lambda(s-\Delta_{\text{F,max}})}{R(\phi_1+\theta_2)} \right.\\ \left. + \int^{s-\Delta_{\text{F,min}}}_{s-\Delta_{\text{F,max}}} \frac{1}{R(\theta_1+\theta_2)} \frac{\partial \lambda(s')}{\partial s'} ds'\right],
\end{multline}

The quantity $\theta$ here denotes the angle measured from point $P$ backward, so $\theta_1 = (\theta-\theta_2$) in $W_{\text{F1}}$. Since $\Delta_{\text{F,max}}(x=0) = R(\phi_1+\theta_2)^3/24 = \Delta_{\text{B}}(\theta=\phi_1+\theta_2)$, the two integrals can be merged into one. Also, since $\Delta_{\text{F,min}}(x=0) = R\theta_2^3/24 = \Delta_{\text{B,max}}(\theta_{\text{max}}=\theta_2)$, the first term in $W_{\text{F1}}$ cancels with the boundary term in $W_{\text{B}}$ (since they have the opposite sign), giving:

\begin{multline}
W(s) = -4ke^2 \left[ \frac{\lambda(s-\Delta_{\text{B}}(\theta=\phi_1+\theta_2))}{R(\phi_1+\theta_2)} \right.
\\+\left. \int^s_{s-\Delta_{\text{B}}(\theta=\phi_1+\theta_2)} \frac{1}{R\theta} \frac{\partial \lambda(s')}{\partial s'} ds'\right].
\end{multline}

As expected, we have recovered case B for the merged magnet with an entrance angle of $(\phi_1+\theta_2)$. This validates our wake expression for case F1, and the first boundary term in Eq.~\eqref{eq:wF1} is necessary in this formulation. Without this term to cancel the boundary term in case B, one could generate free wakes by splitting the magnet into small magnets, which is not physical.

\subsubsection{Case F2}
Similar to E2, the two magnets now bend in the opposite direction. With the same derivation procedure as in case F1, we obtain: 

\begin{multline}
W(s) = -4ke^2 \left[
-\frac{\lambda(s-\Delta_{\text{min}})}{R_2\theta_2} \right.\\+
\frac{(\phi_1-\theta_2)\lambda(s-\Delta_{\text{max}})}{R_1\phi_1^2+2R_2\phi_1\theta_2+2x\phi_1-R_2\theta_2^2}\\ 
\left. +\int^{s-\Delta_{\text{min}}}_{s-\Delta_{\text{max}}} \frac{(\theta_1-\theta_2)}{R_1\theta_1^2+2R_2\theta_1\theta_2+2x\theta_1-R_2\theta_2^2} \frac{\partial \lambda(s')}{\partial s'} ds' \right], 
\label{eq:wF2}
\end{multline}

The result is very similar to case F1, again with $\theta_2 \rightarrow -\theta_2$ and $R_2 \theta_2$ unchanged. 

\subsubsection{Case H1 and H2}
\begin{figure}[!h]
	\centering
	\includegraphics[width=0.4\textwidth]{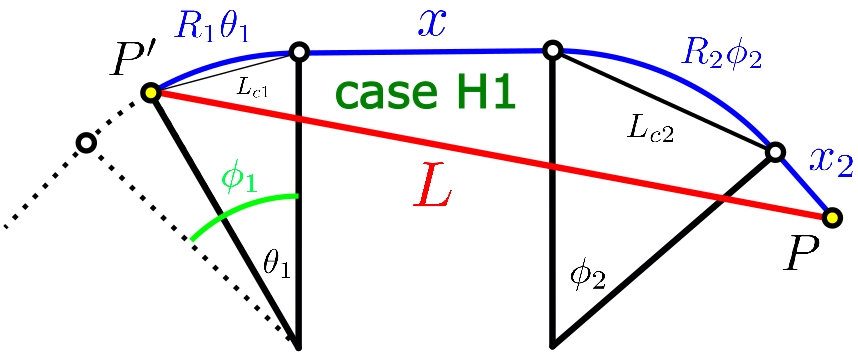}
	\caption{(Color) Geometry for case H1.}
	\label{caseH1}
\end{figure}

Similar to case F, there are two subcases in case H depending on the two bending directions. Fig.~\ref{caseH1} shows the geometry for case H1 in which the two magnets bend in the same direction. The observation point $P$ is located at a distance $x_2$ down the second magnet, just like in case G. 

We follow the same derivation procedure as in case F1 and F2. In the limit of $\gamma^2 \gg L_s/R(\theta_1+\phi_2)^3$ we have:
\begin{multline}
\Delta = \frac{1}{24L_s}[R_1\theta_1^3(R_1\theta_1+4x+4R_2\phi_2+4x_2) 
\\ \pm 6R_1\theta_1^2\phi_2(R_2 \phi_2+ 2x_2) +R_2\phi_2^3(4R_1\theta_1+4x+R_2\phi_2) 
\\ + 4x_2\phi_2^2(3R_1\theta_1+3x+R_2\phi_2)],
\end{multline}
\begin{equation}
w =\frac{32ke^2L_s^2(R_1\theta_1^2 \pm 2R_1 \theta_1\phi_2  \pm 2x\phi_2 \pm R_2\phi_2^2)}{R_1(R_1\theta_1^2+2R_2\theta_1\phi_2+2x\theta_1 \pm R_2\phi_2^2+2x_2(\theta_1 \pm \phi_2))^4},
\end{equation}
\begin{equation}
u(\theta_1) = \frac{4ke^2(\theta_1 \pm \phi_2)}{R_1\theta_1^2+2R_2\theta_1\phi_2+2x\theta_1 \pm R_2\phi_2^2+2x_2(\theta_1 \pm \phi_2)}
\label{eq:uGH}
\end{equation}
Integration by parts gives:
\begin{multline}
W(s) = -\left[-u(0)\lambda(s-\Delta_{\text{min}}) \right.\\
\left.+u(\phi_1)\lambda(s-\Delta_{\text{max}}) + \int^{s-\Delta_{\text{min}}}_{s-\Delta_{\text{max}}} u(\theta_1)\frac{\partial \lambda(s')}{\partial s'} ds'\right].
\label{eq:WH}
\end{multline}

Again, the $\pm$ sign denotes $``+"$ for case H1 and $``-"$ for case H2. Similar to case F, one cannot apply this expression for small $\phi_2$ or large $x$. One can verify that for $x_2 = 0$, we recover the wake expressions for case E1 and E2. 

\section {Theory v.s Simulation}
To test the formulas derived for all cases, we run CSR simulations in Bmad for four different beamlines (details on Bmad simulation and parameter choice are described in Section IV). Each beamline consists of only drifts and bending magnet(s) with lengths described by Table I.   

\begin{table}[h]
\centering
\begin{tabular}{|c|c|c|c|c|c|}
\hline
    Length (mm)& D0 & B1 & D1 & B2 & D2\\\hline
    Beamline A & 60 & 500 & 600 & X & X  \\\hline
    Beamline B & 60 & 133 & 600 & X & X  \\\hline
    Beamline C & 60 & 133 & 70 & 500 & X  \\\hline
    Beamline D & 60 & 133 & 70 & 122 & 600 \\\hline
\end{tabular}
\caption{The length of each element in the four beamlines. In the element names,``D" denotes a drift, and ``B" denotes a bending magnet. Each beamline starts with drift D0, followed by B1, D1, B2, then D2. The symbol ``X" means that the element is absent.}
\label{tab1}
\end{table}

In this section we redefine $s$ to be the longitudinal position of the bunch center from the beginning of the beamline. The old $s$ is replaced by symbol $z$ to avoid confusion. For each beamline we will track a Gaussian bunch ($\sigma_z = 1.078$~mm) and compare the evolution of the wakefield $W(z)$ with the theory. For easy comparison with other literature, we normalize $W(z)$ in all the plots by the characteristic CSR wake \cite{Chris}:
\begin{equation}
W_0 = \frac{kQe}{(R^2\sigma_z^4)^{1/3}},
\label{eq:W0}
\end{equation}
in which $Q$ is the bunch charge, and $N_p = Q/e$ is the number of electrons in the bunch. We choose a small bunch charge of $Q = 1.0$~pC so that the longitudinal distribution $\lambda(z)$ remains Gaussian during the transport. Note that in general $\int \lambda(z) dz = N_p$. The electron energy is chosen to be 42~MeV, which corresponds to $\gamma = 82.2$ for electrons. The two magnets bend in the $opposite$ direction, and the bending radii are $R_1 = 0.808$~m and $R_2 = 0.487$~m. With $R$ set to $R_1$ in Eq.~\eqref{eq:W0}, we have $W_0 = 1.50 \times 10^{-17}$~J/m $=93.7$~eV/m. To see the wake propagation over a reasonable length scale, the final element(s) in each beamline have been made sufficiently long. The length of the first drift D0 does $not$ play a role since any drift before the first magnet is assumed to be infinitely long in the CSR simulation of Bmad.

\subsection{ Beamline A }
The purpose of beamline A and B is to confirm that Bmad simulation agrees with the previously published formulas with one bending magnet (case A,B,C, and D). The magnet in beamline A is made long ($500$~mm) to allow studies of the wake propagation up to the steady state (s-s). As discussed in \cite{Emma}, the s-s occurs when the bunch center is $L_0 = R\theta \gg (24 \sigma_z R^2)^{1/3}$ into the magnet. $L_0$ is called the overtaking distance, and is equal to $257$~mm for magnet B1. The wake in bend B1 has contributions from case A and case B:
\begin{equation}\label{eq:WA1}
W(z) = W_{\text{B,integral}}(z,\theta_1) + W_{\text{B,boundary}}(z,\theta_1) + W_{\text{A}}(z,\theta_1),
\end{equation}
in which $\theta_1=(s - 60$~mm)/$R_1$ is the angle of the observation point into magnet B1. Eq.~\eqref{eq:WA1} is often referred to as the entrance wake. Fig.~\ref{reg1Wtot} below shows the wake propagation in B1 by theory and Bmad simulations, which agree well. The steady state is reached when $s \approx 520$~mm, beyond which the contribution to $W(z)$ comes solely from the $W_{\text{B,integral}}$ term, and becomes independent of $\theta_1$. 

\begin{figure}[!h]
	\centering
	\includegraphics[width=0.5\textwidth]{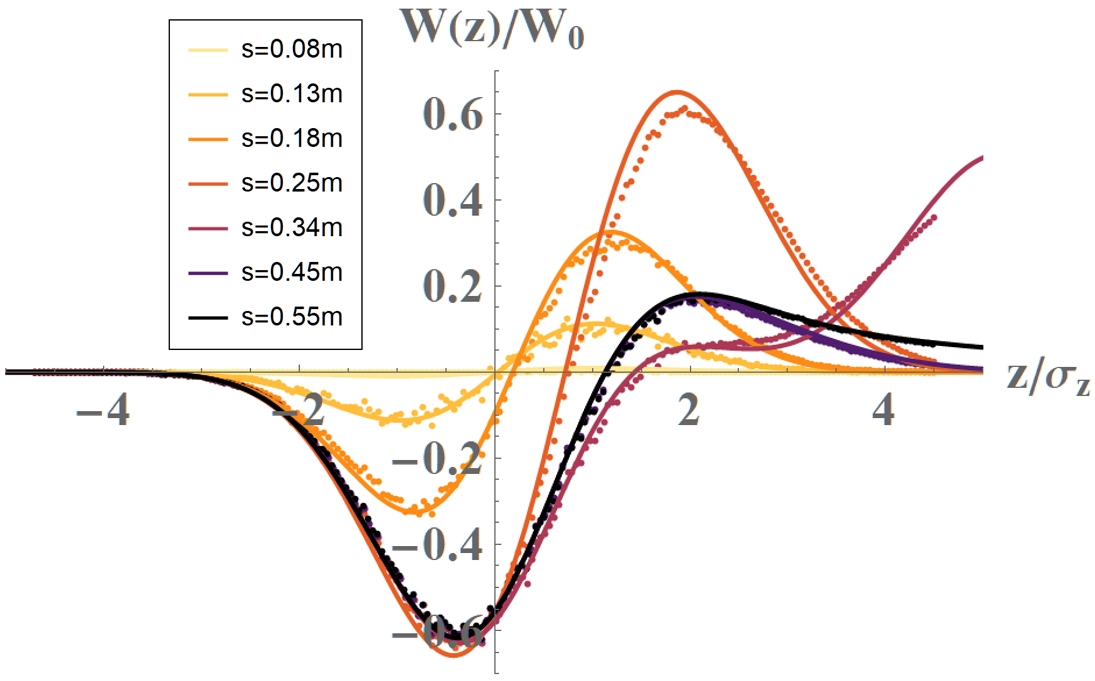}
	\caption{(Color) Evolution of $W(z)$ inside magnet B1 in beamline A. The curves are theory prediction using Eq.~\eqref{eq:WA1}, and the dots are from Bmad simulation. The darkest curve at $s=0.55$~m is the steady state CSR wake.}
	\label{reg1Wtot}
\end{figure}

\begin{figure}[!h]
	\centering
	\includegraphics[width=0.5\textwidth]{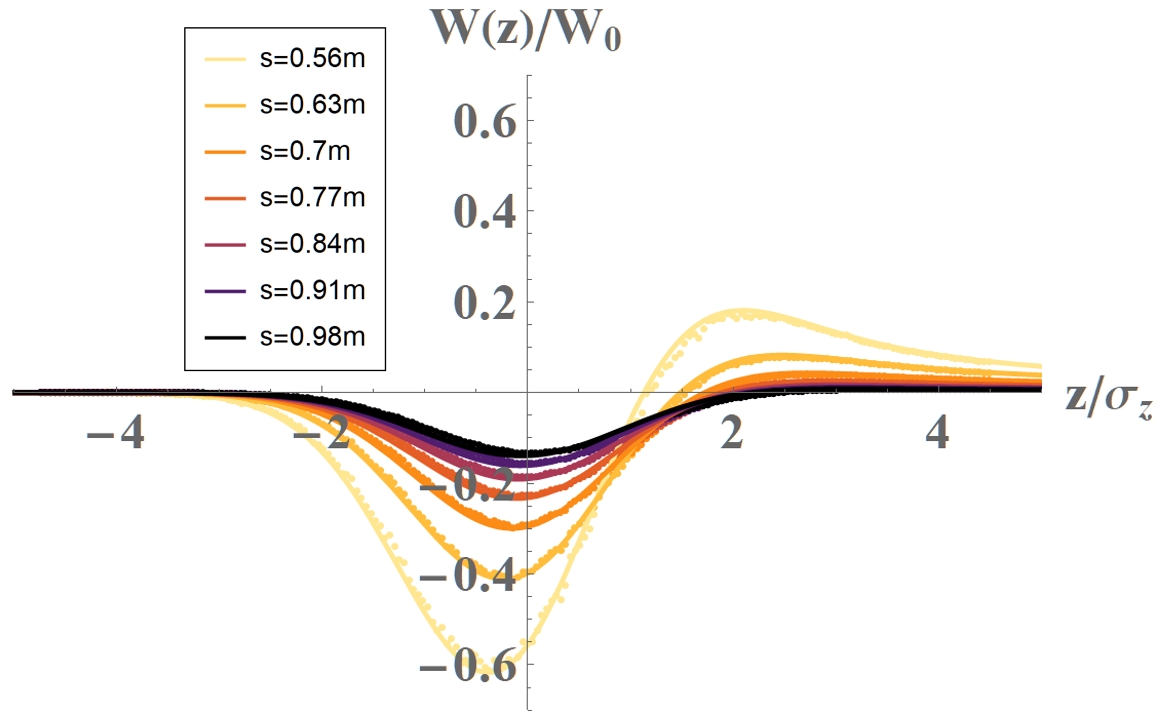}
	\caption{(Color) Evolution of $W(z)$ inside drift D1 in beamline A. The curves are theory prediction using Eq.~\eqref{eq:WA2}, and the dots are from Bmad simulation.}
	\label{reg1Wtot2}
\end{figure}

The wake in drift D1 has contributions from case C and case D:
\begin{equation} 
\begin{aligned}
W(z) &= W_{\text{D,integral}}(z,x,\phi_1)\\
&+W_{\text{D,boundary}}(z,x,\phi_1) + W_{\text{C}}(z, x,\phi_1), \label{eq:WA2}
\end{aligned}
\end{equation}

in which $x = (s - 560~$mm) is the location of the bunch center into the drift D1. Since the s-s is already reached, the only contribution comes from the $W_{\text{D,integral}}$ term. This makes sense since for large $\phi_1$ the other two terms vanish. Fig.~\ref{reg1Wtot2} shows the wake propagation in D1 by theory and Bmad simulations, which again agree well. A similar benchmarking result has been shown in \cite{Sagan}.

\subsection{ Beamline B }
In contrast to beamline A, the magnet's length in beamline B is shorter than the overtaking distance (i.e. $L_m < L_0$), so the s-s cannot be reached. This means all three terms in Eq.~\eqref{eq:WA2} contribute to the exit wake. The wake evolution in D1 is shown in Fig.~\ref{reg2Wtot} below, and we again observe agreement between theory and Bmad simulation.

\begin{figure}[!h]
	\centering
	\includegraphics[width=0.5\textwidth]{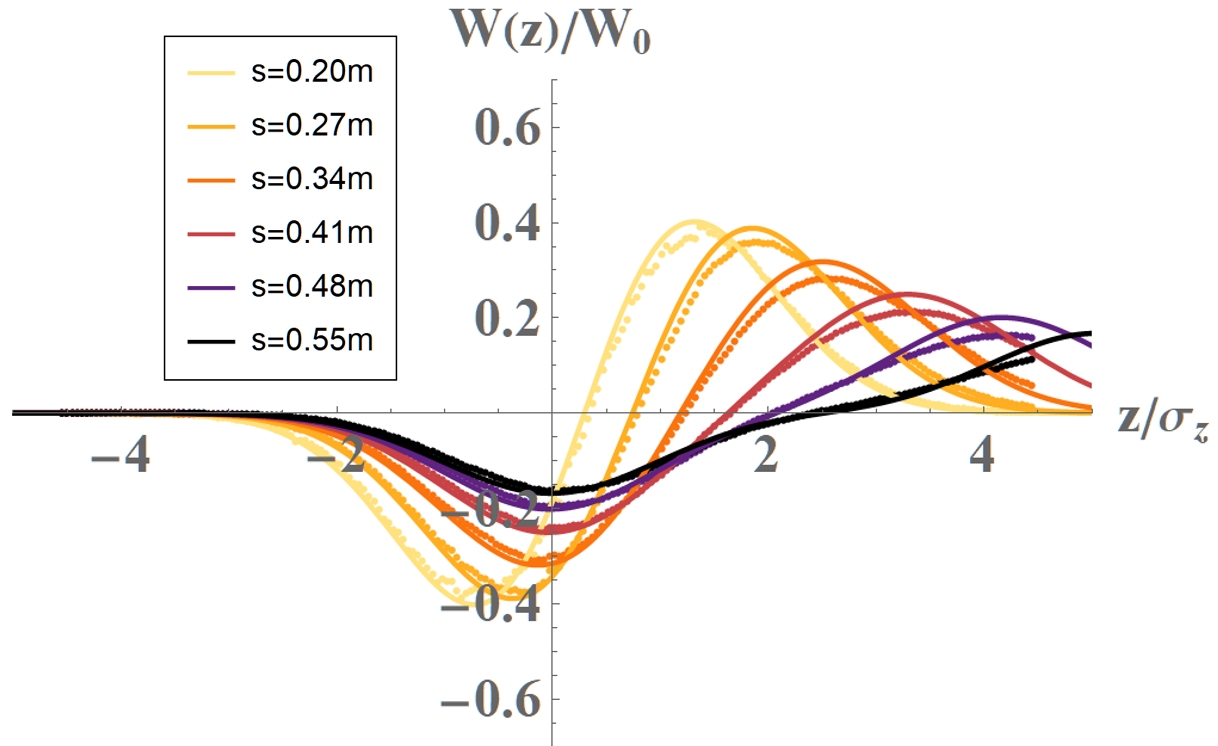}
	\caption{(Color) Evolution of $W(z)$ inside drift D1 in beamline B. The curves are theory prediction using Eq.~\eqref{eq:WA2}, and the dots are from Bmad simulation.}
	\label{reg2Wtot}
\end{figure}

\subsection{ Beamline C }
The purpose of Beamline C is to test the new formulas for case E and F. The wake propagation up to D1 has been shown in beamline B, so here we will focus on $W(z)$ in B2 only. The drift length between the two magnets is made short (70~mm) so that the wake contribution from the first magnet remains significant in the second magnet. The short length also allows us to neglect the wake contribution from the drift itself, because it is the case of small $y_m$ in Eq.~\eqref{eq:wym}.
This neglected contribution corresponds to case A for the second magnet.

The contribution to the total $W(z)$ therefore consists of a total of six terms: 2 term from case B, 3 terms from case F2, and 1 term from case E2. Here we sort the terms into three groups:
\begin{gather}
\begin{aligned}
W(z) &= W_1(z) +W_2(z) +W_3(z),\\
W_1(z) &= W_{\text{B,integral}}(z,\theta_2),\\
W_2(z) &= W_{\text{B,boundary}}(z,\theta_2) + W_{\text{F2,boundary}}^{\text{near}}(z,\theta_2,x),\\ 
W_3(z) &= W_{\text{F2,integral}}(z,\theta_2,x,\phi_1)\\
&+W_{\text{F2,boundary}}^{\text{far}}(z,\theta_2,x,\phi_1)+W_{\text{E2}}(z,\theta_2,x,\phi_1),
\label{eq:WC123}
\end{aligned}
\end{gather}

in which $\theta_2=(s - 263$~mm)/$R_2$ is the angle of the observation point into magnet B2, $x$ is the length of drift D1, and $\phi_1$ is the total bending angle of B1.  $W_1(z)$ has only the integral term of case B for magnet B2. This term is responsible for the steady state CSR wake in magnet B2. $W_2(z)$ has the boundary term of case B and the $near$ boundary term of case F2. Recall that $W_{\text{F2}}$ from Eq.~\eqref{eq:wF2} has two boundary terms. To distinguish them we call the one evaluated at $\theta_1 = \phi_1$ the $near$ term, and the one evaluated at $\theta_1 = \phi_1$ the $far$ term. As discussed in case F, the near term and the boundary term from case B cancel each other if we merge the two magnets (i.e. $W_2(z) = 0$ if $x=0$ and $R_1 = R_2$).
$W_3(z)$ includes the rest of the contribution from magnet B1 and the long drift in front of it. One can identify all six terms in Eq.~\eqref{eq:WC123} based on their names using the tables in Appendix A.

Fig.~\ref{reg3W1}, \ref{reg3W2}, and \ref{reg3W3}, respectively show the evolution of $W_1(z), W_2(z)$, and $W_3(z)$ in magnet B2 ($s>263$~mm) as $\theta_2$ increases. For comparison purposes, the three plots have the same scale, the curves are evaluated at the same longitudinal positions.

\begin{figure}[!h]
	\centering
	\includegraphics[width=0.4\textwidth]{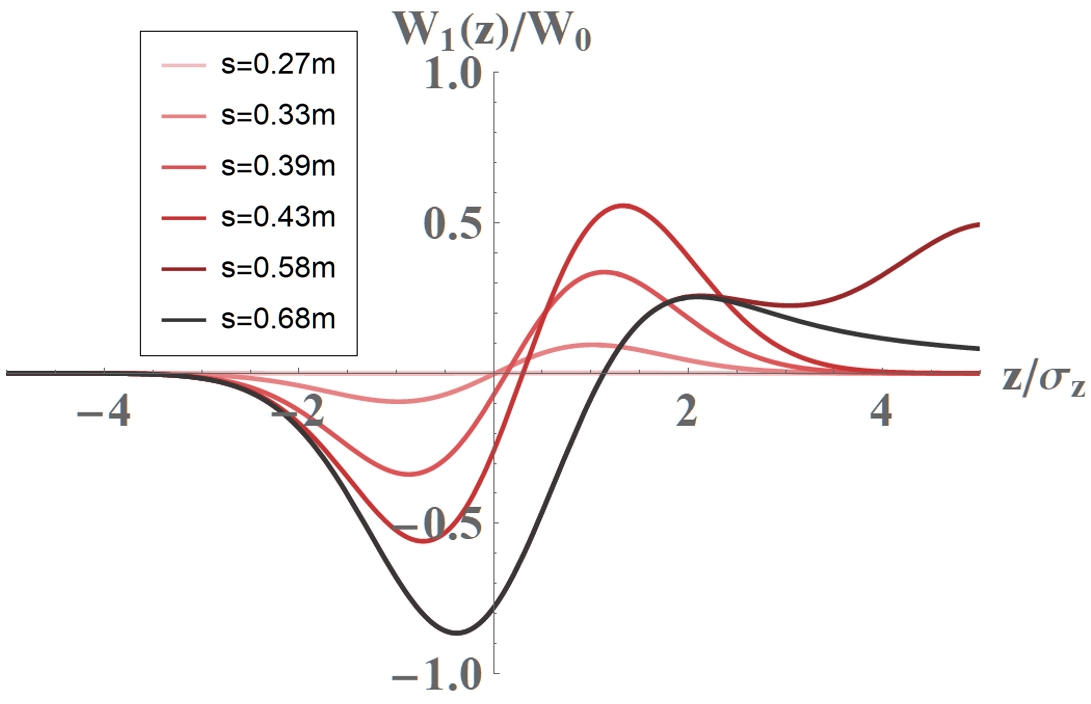}
	\caption{(Color) Evolution of $W_1(z)$ inside magnet B2 in beamline C by theory prediction (See Eq.~\eqref{eq:WC123}.). The evolution is similar to Fig.~\ref{reg1Wtot} since they both show entrance wakes. The darkest curve at $s=0.68$~m corresponds to the steady state wake in B2.}  
	\label{reg3W1}
\end{figure}
\begin{figure}[!h]
	\centering
	\includegraphics[width=0.4\textwidth]{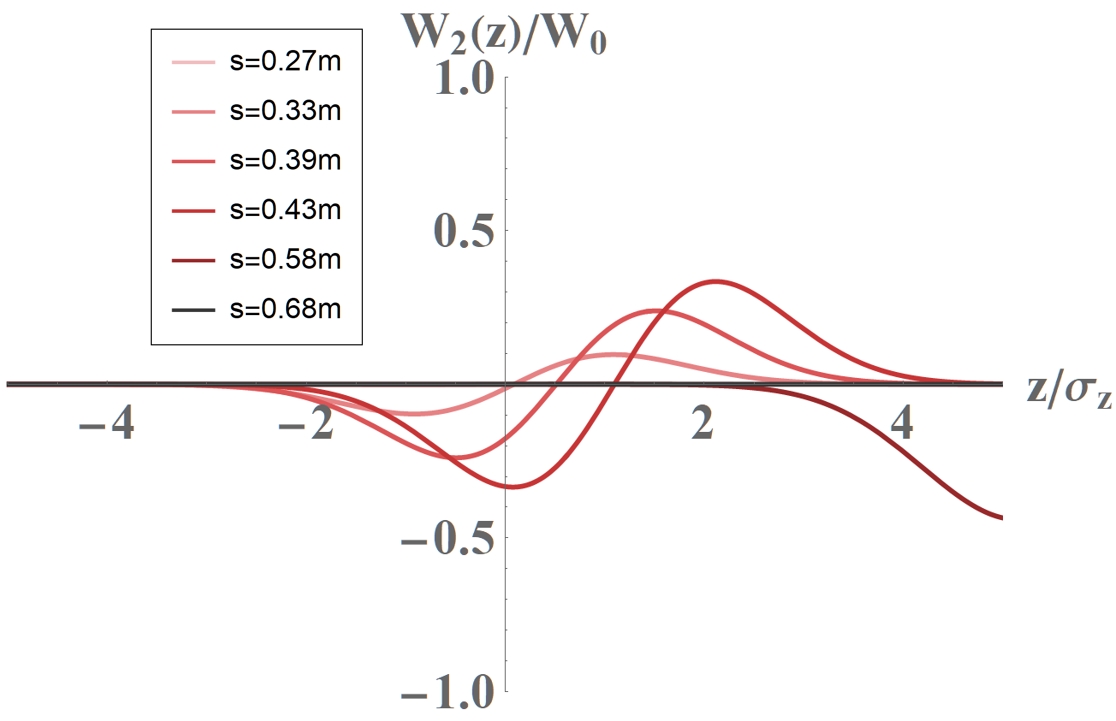}
	\caption{(Color) Evolution of $W_2(z)$ inside magnet B2 in beamline C by theory prediction (See Eq.~\eqref{eq:WC123}.). As $s$ increases the wake ``moves" toward positive $z$. For the darkest curve at $s = 0.68$~m, the wake becomes insignificant for $z<5\sigma_z$.} 
	\label{reg3W2}
\end{figure}
\begin{figure}[!h]
	\centering
	\includegraphics[width=0.4\textwidth]{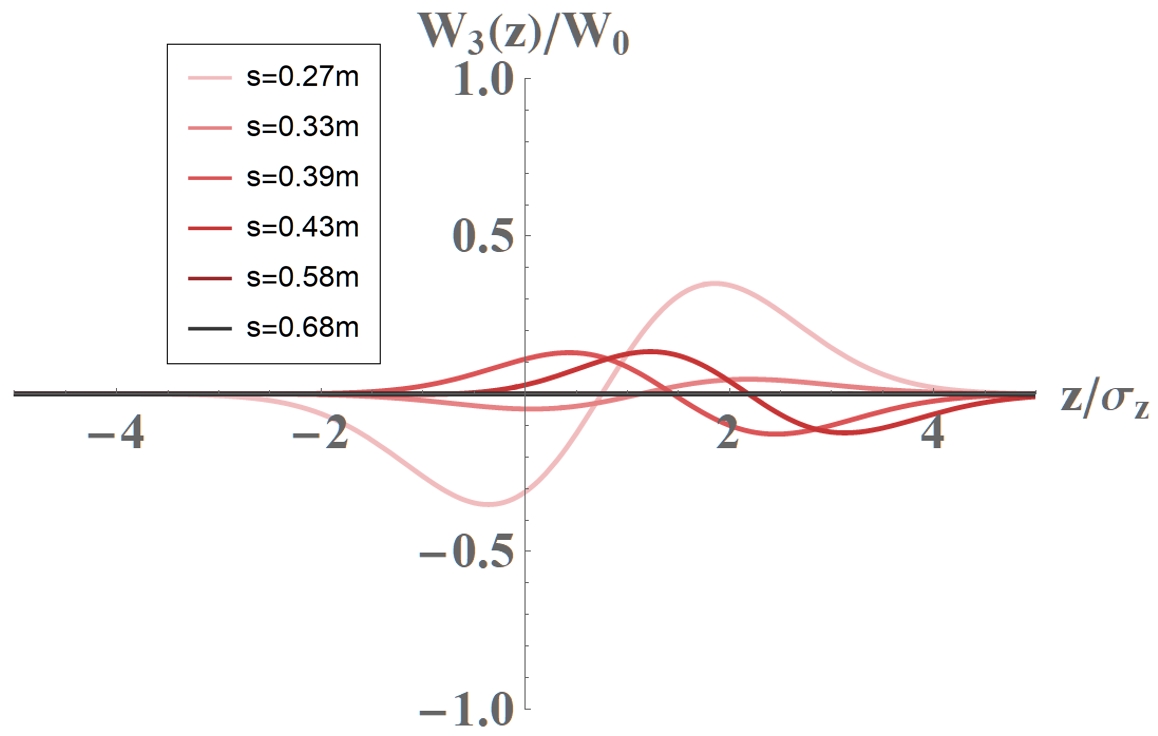}
	\caption{(Color) Evolution of $W_3(z)$ inside magnet B2 in beamline C by theory prediction (See Eq.~\eqref{eq:WC123}.). The wake attenuates for large $s$ since all three terms in $W_3$ come from the first magnet B1.} 
	\label{reg3W3}
\end{figure}

\begin{figure}[!h]
	\centering
	\includegraphics[width=0.5\textwidth]{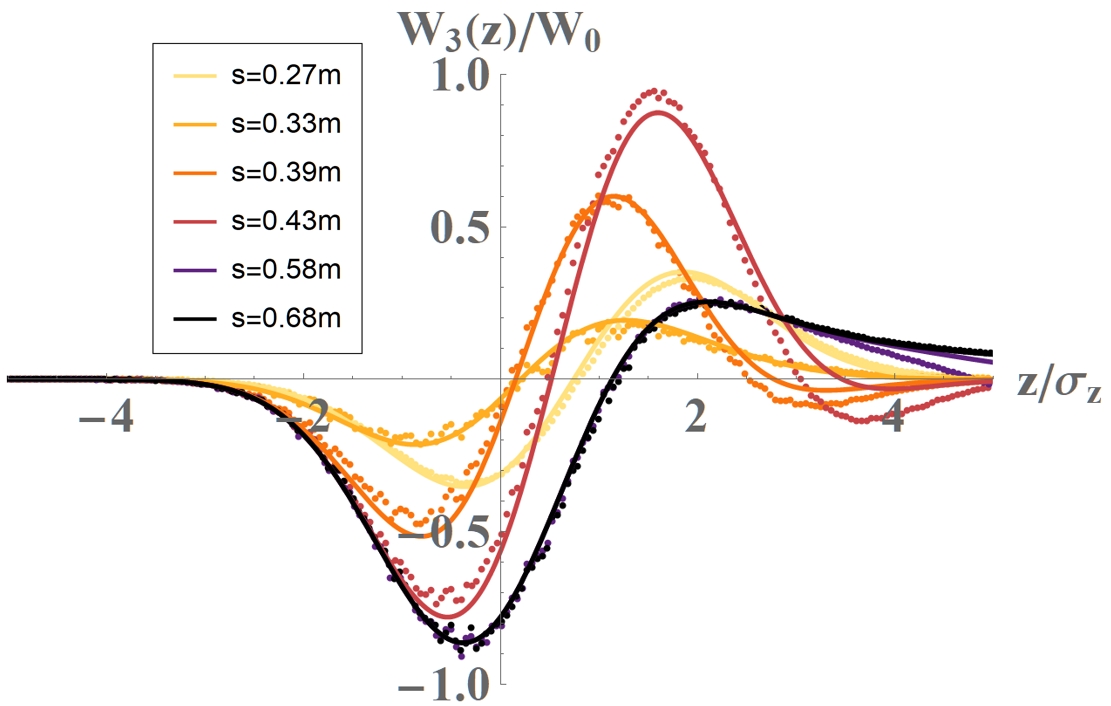}
	\caption{(Color) Evolution of $W(z)$ inside magnet B2 in beamline C. The curves are theory prediction using the sum of three wakes in Eq.~\eqref{eq:WC123}, and the dots are from Bmad simulation.}
	\label{reg3Wtot}
\end{figure}

Fig.~\ref{reg3Wtot} shows the total wake $W(z)$ in B2 and the simulation results from Bmad. As $s$ increases, the main contribution to $W(z)$ shifts from $W_3(z)$ to $W_2(z)$, then eventually to $W_1(z)$. This makes sense since the exit wake from B1 attenuates as the entrance wake from B2 builds up, and finally for $s=0.68$~m the steady state wake is reached, which corresponds to the darkest curve in Fig.~\ref{reg3W1}. Note that the normalization factor $W_0$ in all the plots depends on $R_1$, not $R_2$. This explains why the range of the steady state wake is different in Fig.~\ref{reg1Wtot} and Fig.~\ref{reg3Wtot}.

\subsection{ Beamline D }
\begin{figure}[!h]
	\centering
	\includegraphics[width=0.4\textwidth]{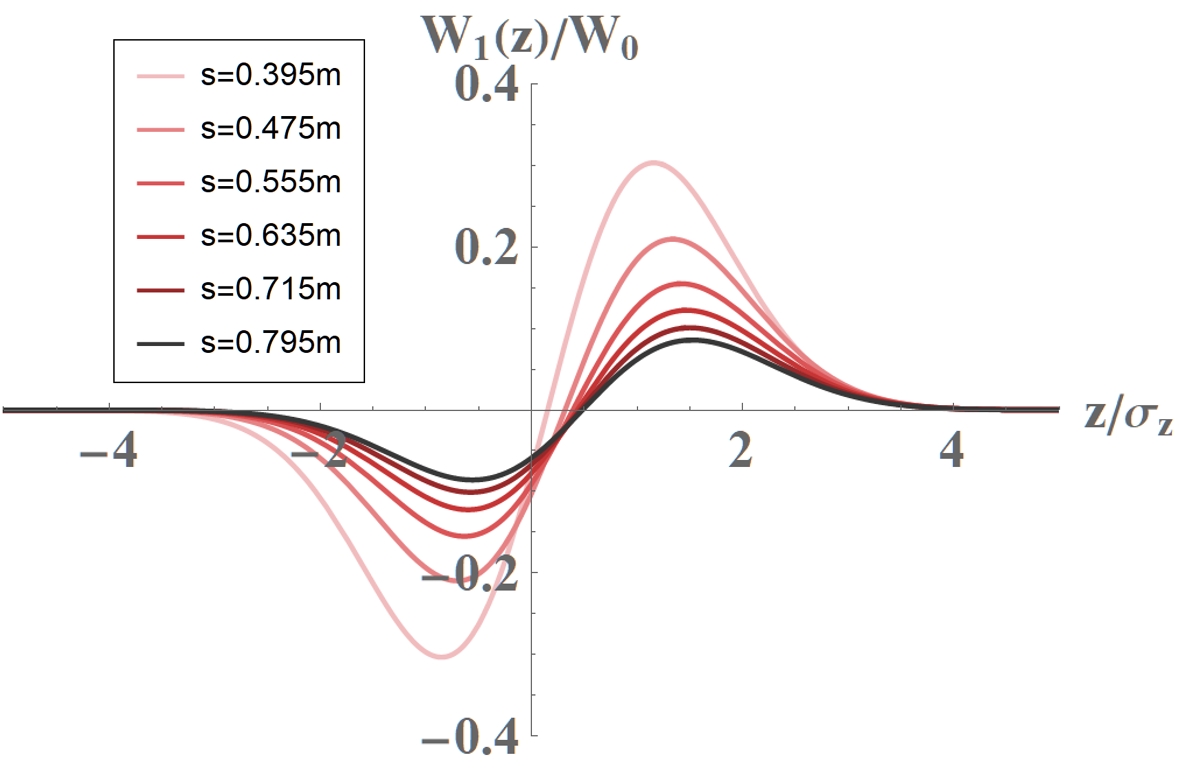}
	\caption{(Color) Evolution of $W_1(z)$ inside drift D2 in beamline D by theory prediction (See Eq.~\eqref{eq:WD123}.). The exit wake attenuates like as expected.}  
	\label{reg4W1}
\end{figure}

\begin{figure}[!h]
	\centering
	\includegraphics[width=0.4\textwidth]{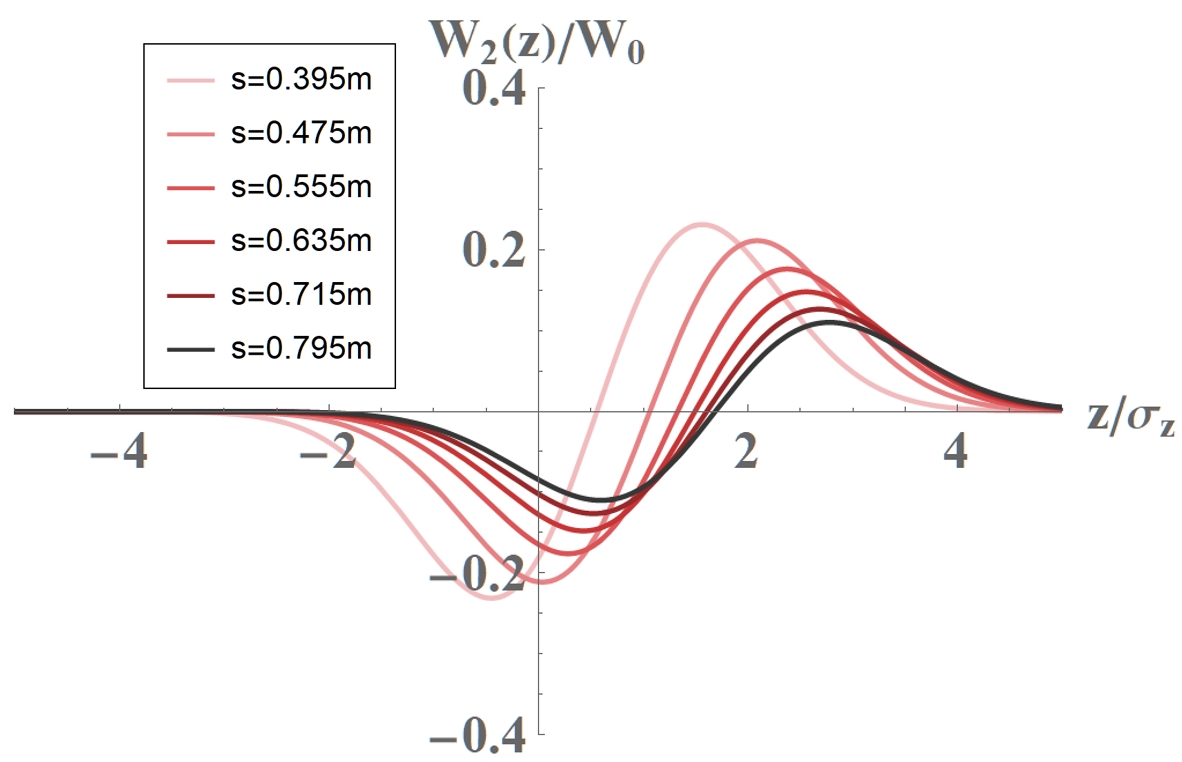}
	\caption{(Color) Evolution of $W_2(z)$ inside drift D2 in beamline D by theory prediction (See Eq.~\eqref{eq:WD123}.). As $s$ increases the wake ``moves" toward positive $z$ and attenuates.} 
	\label{reg4W2}
\end{figure}

\begin{figure}[!h]
	\centering
	\includegraphics[width=0.4\textwidth]{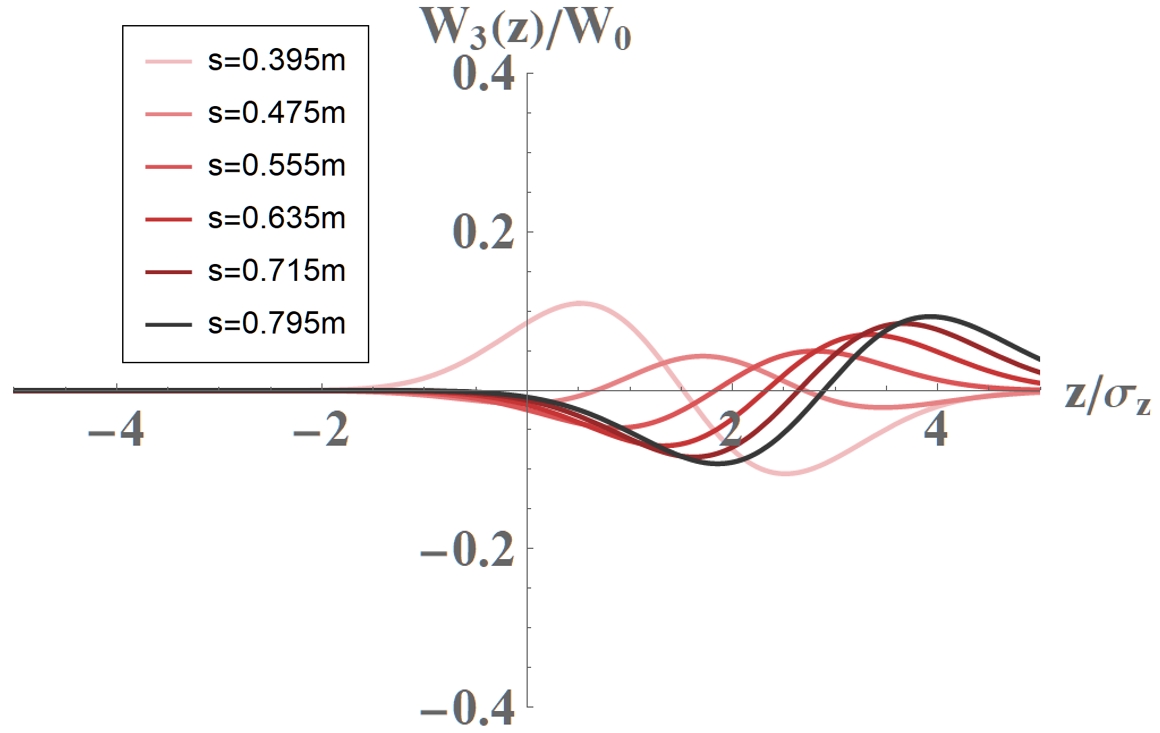}
	\caption{(Color) Evolution of $W_3(z)$ inside drift D2 in beamline D by theory prediction (See Eq.~\eqref{eq:WD123}.). The wake eventually attenuates for large $s$ (not shown). }
	\label{reg4W3}
\end{figure}

The purpose of beamline D is to test the formulas for case G and H. Similar to beamline C, the total wake has contribution from six terms, grouped as: 
\begin{gather}
\begin{aligned}
W(z) &= W_1(z) +W_2(z) +W_3(z),\\
W_1(z) &= W_{\text{D,integral}}(z,x_2,\phi_2),\\
W_2(z) &= W_{\text{D,boundary}}(z,x_2,\phi_2) + W_{\text{H2,boundary}}^{\text{near}}(z,x_2,\phi_2,x),\\
W_3(z) &= W_{\text{H2,integral}}(z,x_2,\phi_2,x,\phi_1)\\+&W_{\text{H2,boundary}}^{\text{far}}(z,x_2,\phi_2,x,\phi_1) +W_{\text{G2}}(z,x_2,\phi_2,x,\phi_1),
\label{eq:WD123}
\end{aligned}
\end{gather}

in which $x_2 = (s - 385~$mm) is the location of the bunch center into the drift D2, and $\phi_2$ is the total bending angle of B2. The six terms are grouped in the same way as in Eq.~\eqref{eq:WC123}, with the following changes in the case names: B $\rightarrow$ D, F2 $\rightarrow$ H2, and E2 $\rightarrow$ G2.
Fig.~\ref{reg4W1}, \ref{reg4W2}, and \ref{reg4W3} respectively show the evolution of $W_1(z), W_2(z)$, and $W_3(z)$ in D2 ($s>385$~mm) as $x_2$ increases. 


\begin{figure}[!h]
	\centering
	\includegraphics[width=0.5\textwidth]{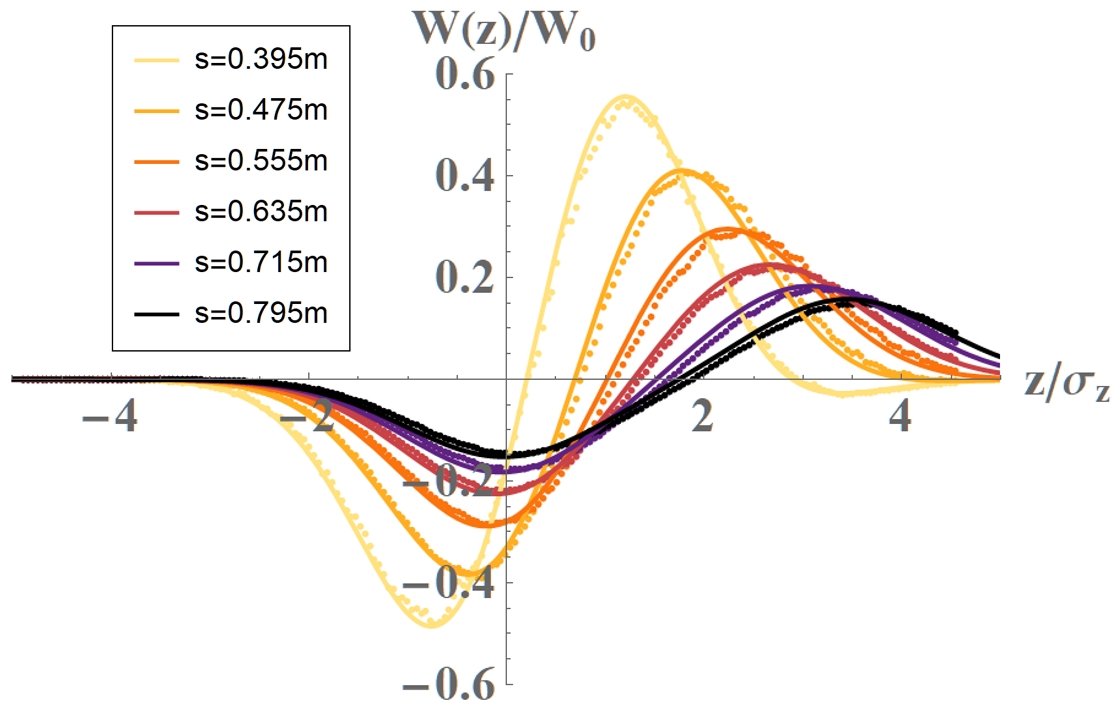}
	\caption{(Color) Evolution of $W(z)$ inside drift D2 in beamline D. The curves are theory prediction using the sum of three wakes in Eq.~\eqref{eq:WD123}, and the dots are from Bmad simulation.}
	\label{reg4Wtot}
\end{figure}

Fig.~\ref{reg4Wtot} shows the total wake $W(z)$ in D2 and the simulation results from Bmad, which agree well. As $s$ increases, the wake amplitude decreases as expected. However, one might note that the amplitude of $W_3(s)$ in Fig.~\ref{reg4W3} does $not$ monotonically decrease for all $s$. This occurs because the denominator of $u(\theta_1)$ in Eq.~$\eqref{eq:uGH}$ (and therefore $u(\phi_1)$ in Eq.~$\eqref{eq:WG}$ and Eq.~$\eqref{eq:WH}$) changes signs as $x_2$ increases. This can only happen when the two magnets bend in the opposite direction, so that the denominator is not always positive and increasing with $x_2$.

\subsection{Discussion}

Beamline A and B show that Bmad simulation agrees with the existing formulas for $W(z)$ with one bending magnet. Moreover, Beamline C and D show how to apply the new formulas for a system with two bending magnets, and agreements with simulation results support that these formulas work. Note that the simplified formulas might give inaccurate results if $\gamma$ or $\theta$ ( the observation angle into either magnet ) is too small, or if $x$ ( the drift length between the two magnets ) is too large. In that case the un-approximated formulas have to be used. While CSR simulation in Elegant has implemented the formulas with one bending magnet, Bmad uses $none$ of these approximated formulas (See Section IV for details.).

In addition, one can use these formulas to calculate the energy loss $E_{\text{loss}} = \int W(z) \lambda(z) dz$ and the increase in energy spread due to CSR, as long as $\lambda(z)$ remains unchanged. In reality the longitudinal distribution $\lambda(z)$ might vary significantly due to a high bunch charge or transverse particle motions. This extension of CSR theory from one bend to two bends is essential for short-bend accelerators like CBETA. For more extreme systems with CSR extending over three or more bends, no approximated expressions have been derived using the Lienard-Wiechert formula, and numerical simulations are recommended. Alternatively, an exact 1D model using the Jefimenko's form of Maxwell's equations has been derived in \cite{Chris}. 

\section{Bmad CSR Simulation Overview}

\begin{figure}[!htb]
   \centering
   \includegraphics*[width=230pt]{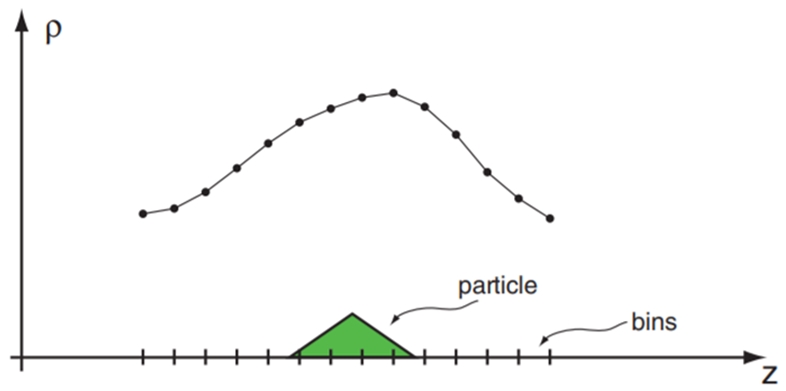}
   \caption{Bmad implementation of CSR. The bunch is divided into $N_b$ bins in the longitudinal direction for calculation of CSR kicks.}
   \label{Bmad_CSR}
\end{figure}

Cornell University has developed a simulation software called Bmad to model relativistic beam dynamics in customized accelerator lattices \cite{bmad}, and subroutines have been established to include CSR calculations \cite{Sagan}. Fig.~\ref{Bmad_CSR} shows how Bmad divides a bunch of particles into a number of bins ($N_b$) in the longitudinal direction. During beam tracking, $N_b$ is constant, and the bin width is dynamically adjusted at each time step to cover the entire bunch length. The contribution of a macro-particle to a bin’s total charge is determined by the overlap of the particle’s triangular charge distribution and the bin. With $\Delta z_b$ denoting the bin width and $\rho_i$ denoting the total charge in the $i^{\text{th}}$ bin, the charge density ($\lambda_i$) at the bin center is taken to be $\rho_i/\Delta z_b$. In between the bin centers, the charge density is assumed to vary linearly.

With integration by parts the CSR wake seen by the bunch can be written as:
\begin{eqnarray}
W(z) = \int_{-\infty}^{\infty} dz' \frac{d\lambda(z')}{dz'}I_{\text{CSR}}(z-z'),
\end{eqnarray}
in which 
\begin{equation}
I_{\text{CSR}}(z-z') = - \int^z_{-\infty}w(z-z'')dz''.
\end{equation}

The energy kick $\Delta\mathcal{E}$ received by a particle centered at the $j^{\text{th}}$ bin, after travelling for a distance $\Delta s$, is therefore modelled in Bmad as \cite{Sagan}:
\begin{gather}
\Delta\mathcal{E} = \Delta s \sum_{i=1}^{N_b}(\lambda_i-\lambda_{i-1})\frac{\hat I_{\text{CSR}}(j-i)+\hat I_{\text{CSR}}(j-i+1)}{2},
\end{gather}

in which $\hat I_{\text{CSR}}(j) \equiv I_{\text{CSR}}(z=j\Delta z_b)$.

CSR simulation results in Bmad have been benchmarked with CSR theory and other simulation codes including A$\&$Y and elegant \cite{Sagan}. Additional benchmarking with a system of two magnets have been shown in Section III of this paper. Bmad CSR simulation also allows users to include the space charge calculation for high energy and the one dimensional vacuum shielding effect. In 2017 the Bmad library was further developed, which can now handle the case when the design orbit of the beam does $not$ follow the reference orbit of the lattice \cite{Chris_IPAC}. This is exactly the case for the FFA beamline in CBETA, which consists of displaced quadrupole magnets. 

Given a bunch with fixed charge $Q$, the two most important parameters in CSR simulations are the number of particles ($N_p$) and bins ($N_b$). A large $N_p$ generally increases the simulation accuracy at the cost of computation time. It is usually recommended to have $Np \geq$ 100k, but a beamline with more or longer curved trajectories may require more. Choosing $N_b$ is not as straightforward as $N_p$. A small $N_b$ can result in inaccurate calculation of CSR kicks due to low longitudinal resolution. However, if $N_b$ is too large, the number of particles per bin might become too small, resulting in numerical noise. A proper choice of $N_b$ therefore depends heavily on $N_p$, the bunch parameters, and the lattice itself. For the four beamlines in Section III we have chosen $N_p = 400k$ and $N_b=200$. For a large lattice like CBETA, convergence tests are recommended to produce convincing results (See section V subsection C). 

\section{CBETA Simulation Results}

\begin{figure}[!htb]
   \centering
   \includegraphics*[width=235pt]{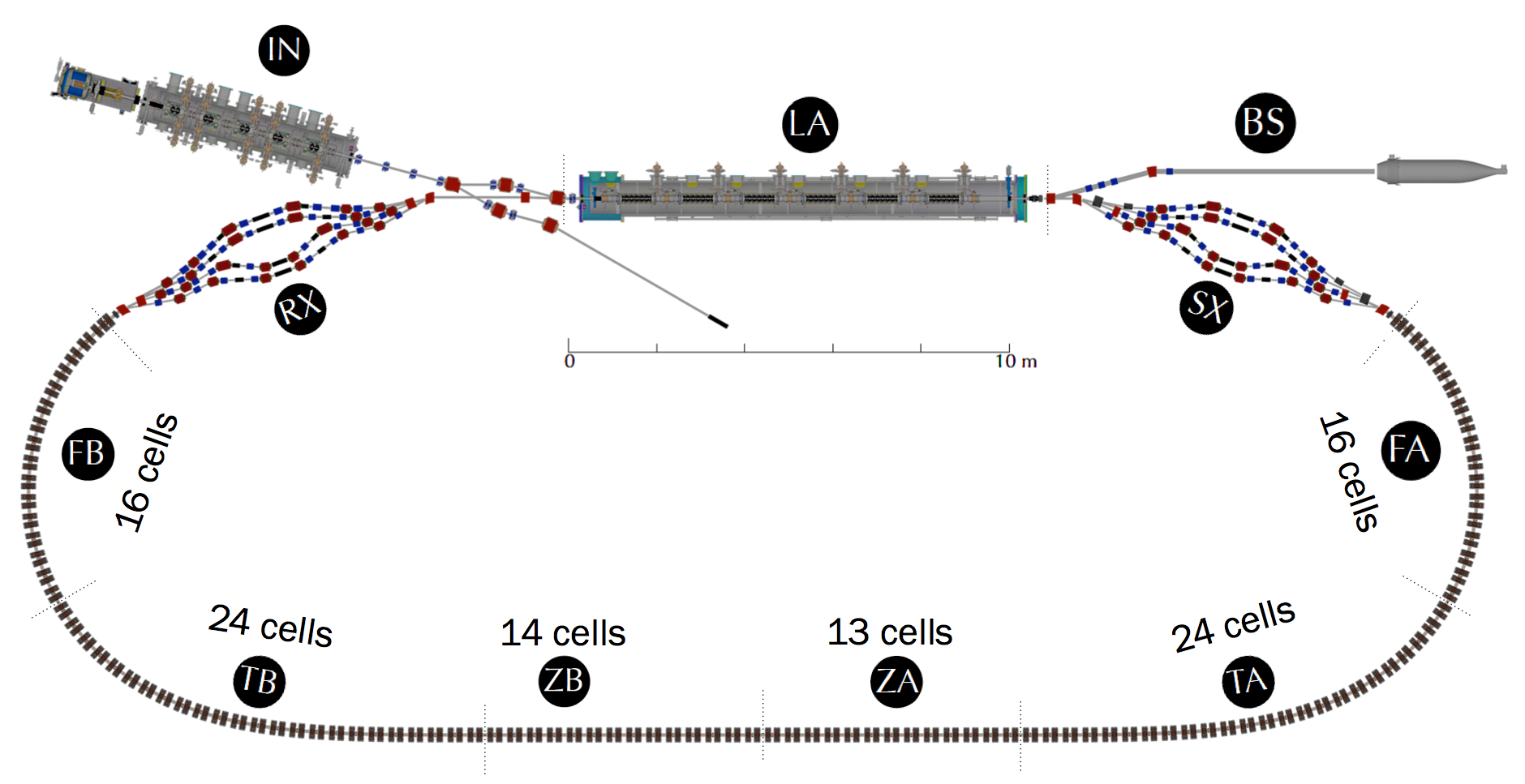}
   \caption{Layout of CBETA. The sections labeled (IN) and (LA) are the injector and MLC cavities respectively. Sections (FA), (TA), (ZA), (ZB), (TB), and (FB) form the FFA beamline which accommodate all recirculating orbits with design energies ranging from 42 MeV to 150 MeV. Sections (SX) and (RX) are splitters and recombiners which control the path-length of each recirculation pass.}
   \label{CBETA}
\end{figure}

CBETA has been constructed at Cornell University's Wilson Laboratory. As a collaboration with BNL, CBETA is the first multipass ERL with a Fixed Field Alternating (FFA) lattice. Four turn energy recovery was first achieved in December 2019. It also serves as a prototype accelerator for electron coolers of Electron Ion Colliders (EICs). The EIC project in the US will benefit from this new accelerator \cite{ipac2017}. Fig.~\ref{CBETA} shows the design layout of CBETA, which is a 4-turn ERL with maximum electron beam energy of 150 MeV. This energy is achieved by first accelerating the electron beam to 6~MeV by the injector (IN). The beam is then accelerated by the Main Linac Cryomodule (MLC) cavities (LA) four times to reach 150~MeV, then the beam is decelerated four times down to 6~MeV before being stopped (BS). The beam passes through the MLC cavities for a total of eight times, each time with an energy gain of $\pm$36~MeV. Field energy in the cavities is transferred to the beam during acceleration, and is recovered during deceleration. Transition from acceleration to deceleration is achieved by adjusting the path-length of the forth recirculation turn to be an odd multiple of half of the RF wavelength. The path-length of all the other turns is an integer multiple of the RF wavelength. CBETA can also operate as a 3-turn, 2-turn, or 1-turn ERL with properly adjusted configuration.

\begin{figure}[!htb]
   \centering
   \includegraphics*[width=235pt]{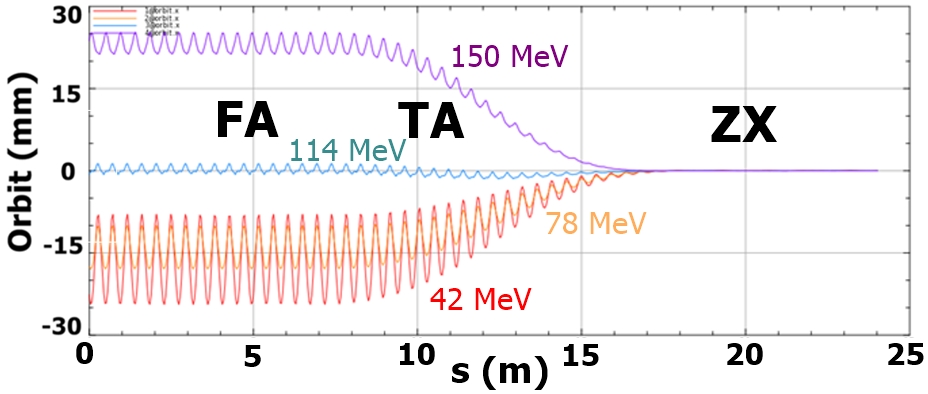}
   \caption{(Color) The orbits of the four design energies inside the first half of the FFA beamline, including FA, TA, and ZA. The second half (not included) has approximately mirrored orbits and optics.}
   \label{FFA}
\end{figure}

Fig.~\ref{FFA} shows the orbits of the four design energies inside the first half of the FFA beamline. Section FA consists of 16 periodic cells, and the periodicity in orbits is broken at the transitional TA section. All four orbits reach zero value at the straight ZA section. The entire FFA beamline consists of permanent Halbach magnets with primarily dipole and quadrupole components in the good field region \cite{optics_ipac2018}. Because the four orbits have different horizontal offsets, they see different equivalent dipole strengths. The main contribution of the CSR effects in CBETA comes from the FA and FB sections, in which the bunches undergo the most curved trajectories, especially at 42~MeV. The beamline D in Section III, with the length of drift D2 shortened to 70~mm, corresponds to one equivalent FA cell seen by the 42~MeV orbit. This shows that the extended theory in Section II with two bending magnets is essential for CBETA. Rather than the traditional theory with only one magnet, the new theory has to be used whenever magnets are similarly close.

Subsections A and B below show the CSR results with the CBETA 1-turn and 4-turn mode for various bunch charges $Q$. The initial bunch distribution has been pre-simulated using GPT tracking up to the end of the LINAC pass 1 (42~MeV) to account for the space charge effect at low energy \cite{cbetacdr}. The GPT beam has a bunch length of 4.0~ps. The CSR parameters are chosen to be $N_p = 10^6$ and $N_b = 2500$. Note that some of these results have been presented in \cite{NAPAC2019}.

\subsection{CBETA 1-turn results}

\begin{figure}[!htb]
   \centering
   \includegraphics*[width=235pt]{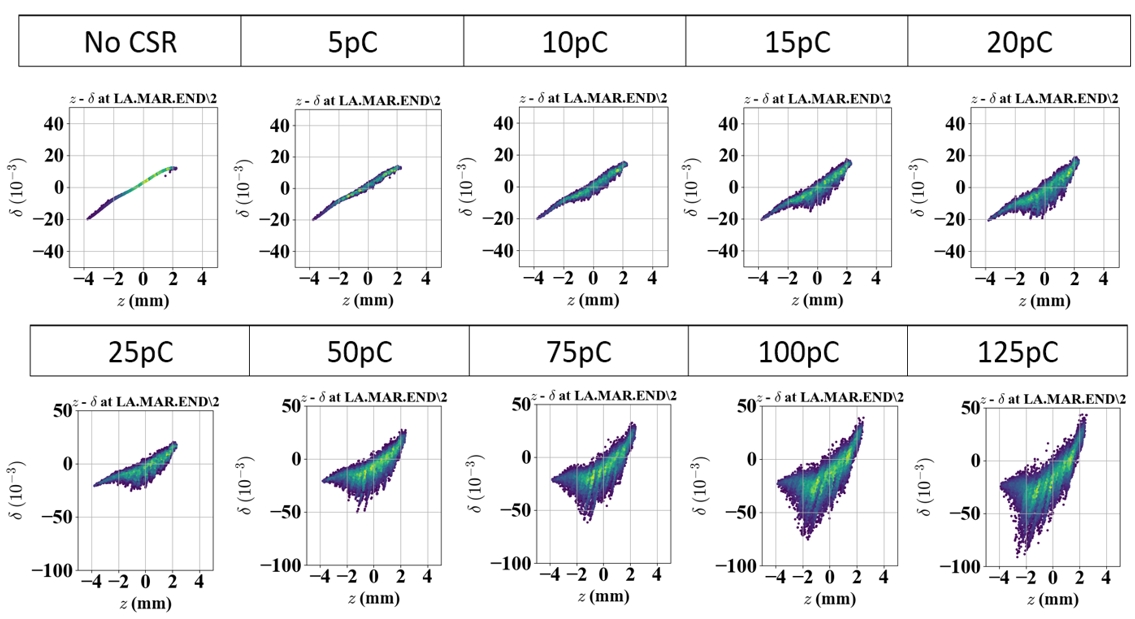}
   \caption{The $z-\delta$ distribution after each of the 8 LINAC passes for CBETA 1-turn with various $Q$.}
   \label{1pf1}
\end{figure}

Fig.~\ref{1pf1} shows the longitudinal phase space distributions of the tracked bunch at the end of LINAC pass 2, where the beam  has returned to 6~MeV, for different $Q$. As $Q$ increases, the CSR effect becomes more significant, causing the increase in energy spread and, via lattice dispersion, the increase in horizontal beam emittance. At $Q =$ 50~pC, 50 out of $10^6$ particles have relative energy spread exceeding $\pm 5\%$. The ideal energy acceptance of the CBETA beam stop is, assuming no halo and other undesired effects, $\pm 7\%$. This limit is exceeded by at least one particle for a bunch charge between 75~pC to 100~pC. This result indicates that CBETA's 1-turn lattice can operate with a 75~pC bunch without particle loss due to CSR. With the maximum repetition rate of 1.3~GHz, this corresponds to a beam current of 97.5~mA, well exceeding the high design current of 40~mA. At $Q =$ 125~pC, which corresponds to 160~mA, 1300 out of $10^6$ particles are lost due to excessive energy spread. 

\subsection{CBETA 4-turn results}

\begin{figure*}[!htb]
   \centering
   \includegraphics*[width=500pt]{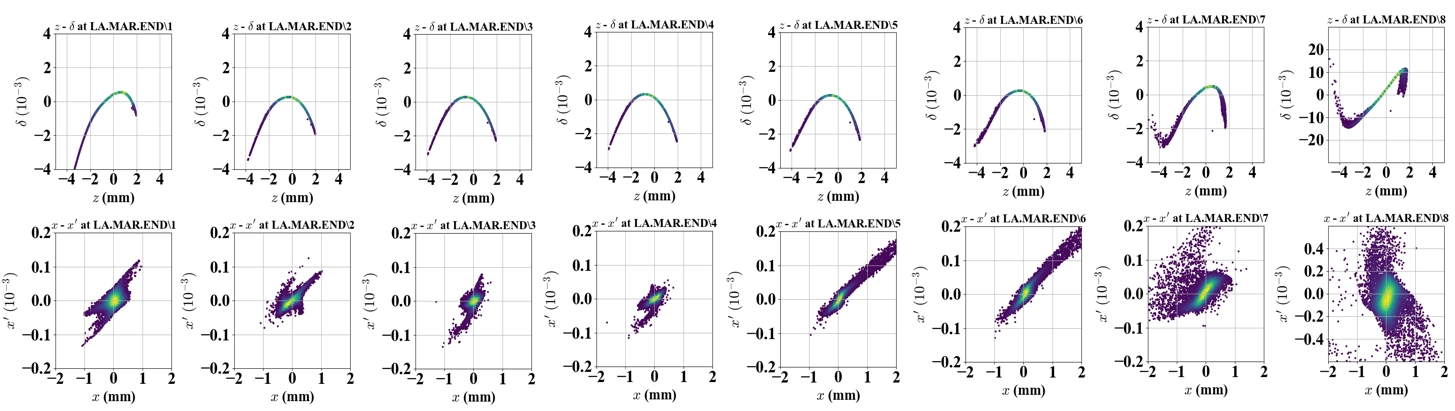}
   \caption{The $x-x'$ and $z-\delta$ distributions after each of the 8 LINAC passes for CBETA 4-turn with no CSR.}
   \label{4pf1}
\end{figure*}
\begin{figure*}[!htb]
   \centering
   \includegraphics*[width=500pt]{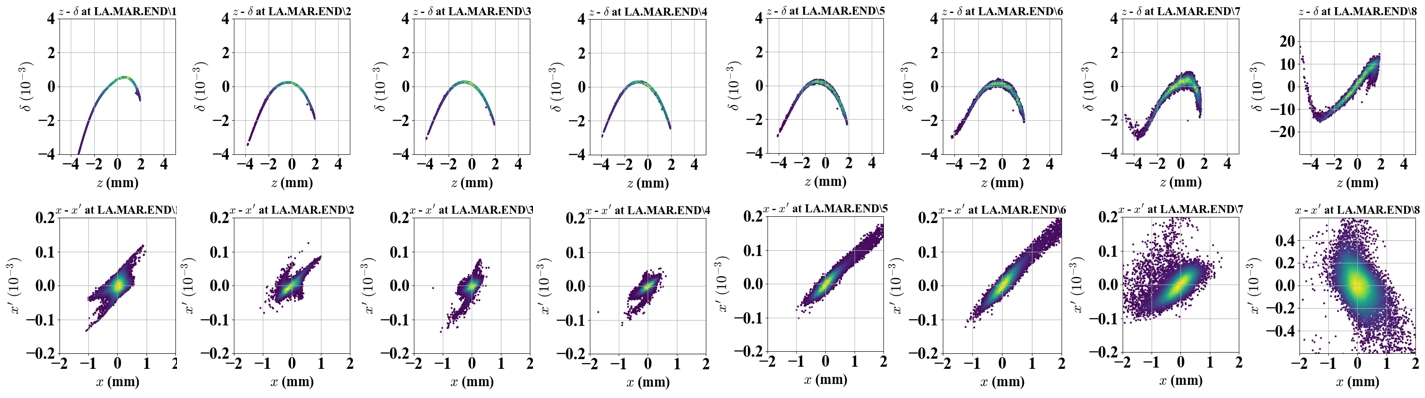}
   \caption{The $x-x'$ and $z-\delta$ distributions after each of the 8 LINAC passes for CBETA 4-turn with $Q=$ 1~pC.}
   \label{4pf2}
\end{figure*}
\begin{figure*}[!htb]
   \centering
   \includegraphics*[width=500pt]{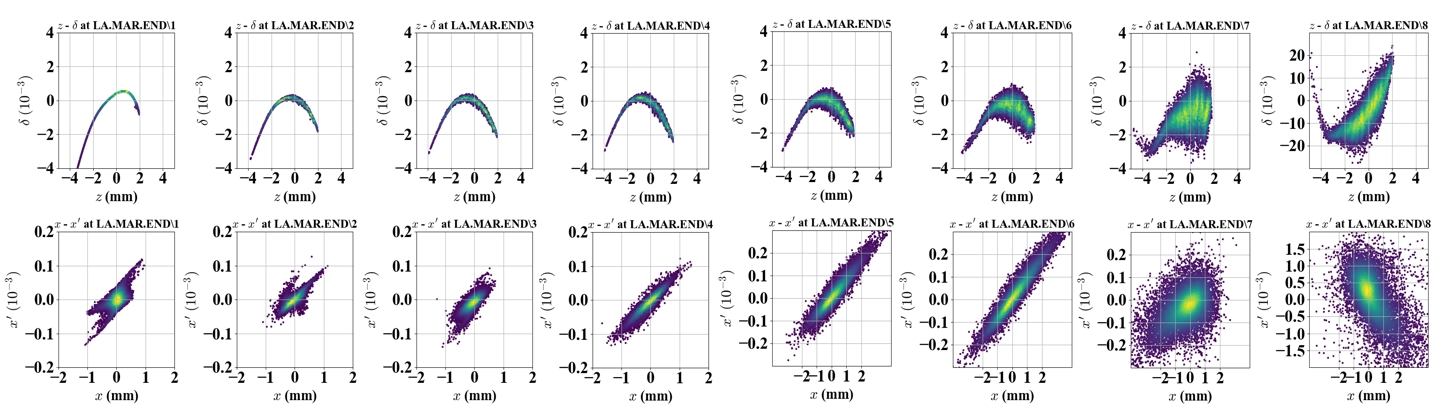}
   \caption{The $x-x'$ and $z-\delta$ distributions after each of the 8 LINAC passes for CBETA 4-turn with $Q=$ 5~pC.}
   \label{4pf3}
\end{figure*}

Fig.~\ref{4pf1} shows the longitudinal and horizontal phase space distributions of the tracked bunch at the end of each LINAC pass, from 1 to 8, with no CSR effects. Fig.~\ref{4pf2} and Fig.~\ref{4pf3} show the corresponding results with CSR effects, for $Q=$ 1~pC and 5~pC respectively. As observed in the 1-turn results, both the energy spread and beam emittance increase as $Q$ increases. Moreover, the energy spread builds up over the recirculation passes. Note that both $x'$ and $\delta$ are dimensionless quantities (scaled by the reference momentum of each pass), which explains why the spreads increase more severely during the four decelerating passes than the four accelerating passes. 

For $Q=$ 1~pC, 215 out of a $10^6$ particles have been lost during the final two decelerating passes. However, all the surviving particles have a final energy spread of less than $\pm 5\%$, which is acceptable for the beam stop. These results show promise for the 4-turn machine to reach its design current of 1~mA, which requires a bunch with $Q \geq$ 3~pC at the maximum repetition rate of 325~MHz. During the decelerating passes, micro-bunching structures can be seen in the longitudinal phase space, e.g. in the four top-right plots of Fig.~27. To what extent this is a physical or a numerical effect requires further study.

\subsection{Convergence Test}
\begin{figure}[!htb]
   \centering
   \includegraphics*[width=0.4\textwidth]{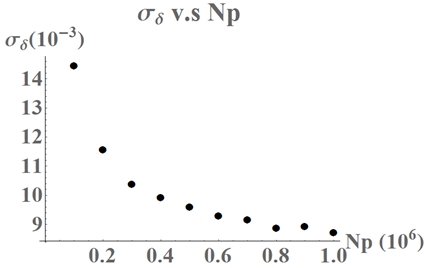}
   \caption{The final rms energy spread for CBETA's 4-turn CSR simulation with varying $N_p$. $N_p$ ranges from $10^3$ to $10^6$, and $N_b$ is fixed at 2500.}
   \label{converge}
\end{figure}

As discussed in Section IV, CSR simulation results can vary significantly based on the choice of numerical parameters, especially for a large lattice like CBETA. Fig.~\ref{converge} shows how the final rms energy spread $\sigma_{\delta}$ varies with $N_p$ ($N_b = 2500$ fixed) for CBETA's 4-turn lattice. We see that $\sigma_{\delta}$ converges from more than 0.014 at $N_p= 100k$ to about 0.009 for $N_p > 800k$. This particular convergence test indicates that with $N_b = 2500$, $N_p > 800k$ is required to produce legitimate results for CBETA's 4-turn. In general there can be multiple CSR parameters, and multi-dimensional convergence tests may be required to produce the most physically representative results. For more examples, see \cite{Tsai}.

\subsection{Mitigation and Future Plan}
\begin{figure}[h!]
   \centering
   \includegraphics*[width=0.35\textwidth]{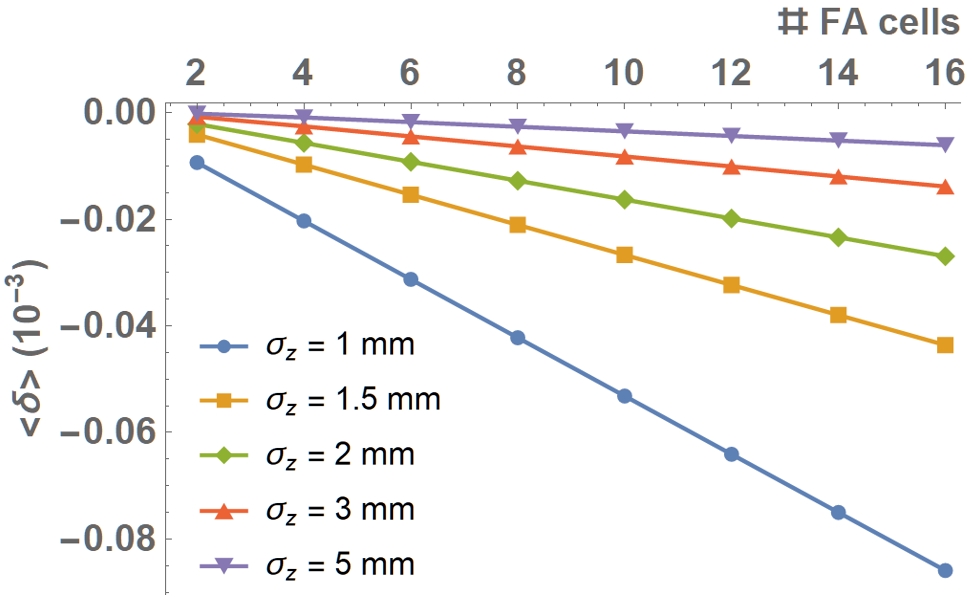}
   \caption{The relative energy loss $<\delta>$ of a Gaussian beam with various initial bunch lengths $\sigma_z$. }
   \label{16FA_mu}
\end{figure}

\begin{figure}[h!]
   \centering
   \includegraphics*[width=0.35\textwidth]{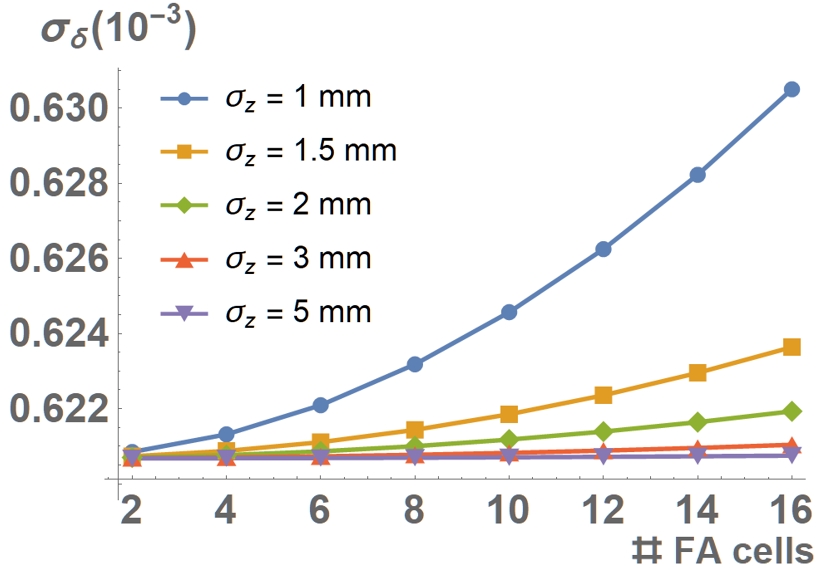}
   \caption{The relative energy spread $\sigma_{\delta}$ of a Gaussian beam with various initial bunch lengths $\sigma_z$. }
   \label{16FA_delta}
\end{figure}

Two methods have been proposed to mitigate the CSR effect. The first method is to increase the bunch length beyond the 4.0~ps used here. Fig.~\ref{16FA_mu} and Fig.~\ref{16FA_delta} below respectively show the relative energy loss $<\delta>$ and energy spread $\sigma_{\delta}$ due to CSR as a 25~pC Gaussian beam traverses the 16 FA cells along the 42~MeV orbit with various initial bunch lengths $\sigma_z$. As expected by theory, both the energy loss and spread decrease as $\sigma_z$ increases. 

Although increasing $\sigma_z$ can mitigate CSR effects in the FFA beamline, it introduces greater energy spread at the LINAC and can lead to undesired ERL operation. Further simulations are required to calculate the optimal $\sigma_z$ for CBETA.

The second method is to include metal shielding, which behaves like waveguides preventing the propagation of CSR fields below the cutoff frequency. Theory and existing experimental data have shown that shielding can potentially suppress energy loss and energy spread of the bunch \cite{Chris}\cite{BNL_shield}\cite{Tsukasa}. While all the simulation results in this paper have assumed CSR propagation in free space, Bmad already has the shielding effect implemented with the method of image charges \cite{bmad}.

The numerical challenge of this method is twofold, it increases the computation time, and convergence tests become more complicated due to an increased number of parameters. The main practical challenge of this method is that CBETA's vacuum chamber height would have to be significantly reduced, making vertical beam steering more challenging. Simulations with shielding for the FFA beamline are currently in progress.

\section{CONCLUSION}
The ultra-relativistic CSR wake expressions with one bending magnet formulated in \cite{Saldin} and \cite{Emma} have been rederived and extended to a system with two bending magnets, and can now be applied for the case when the wake from the first magnet leaks into the second magnet. This can occur to beamlines with magnets placed close to each other, such as the FFA beamline of CBETA. We show that the new terms for wake leakage from one magnet to the next are very relevant. The derived formulas have been compared with Bmad CSR simulation, and agreements have been observed. 

Bmad CSR simulations have been run for the CBETA 1-turn and 4-turn lattice, and increase in energy spread with the bunch charge and the number of recirculation passes have been seen. While the target beam currents are shown to be achievable, more convergence tests on the CSR numerical parameters are required to study the potential microbunching effects. Mitigation to the CSR effects including increasing the bunch length and introducing metal shielding have been proposed, and further simulations are currently in progress.

\section{ACKNOWLEDGMENTS}
We acknowledge Christopher Mayes for his academic knowledge on CSR theories and David Sagan for assistance with Bmad simulations. This work was performed with the support of NYSERDA (New York State Energy Research and Development Agency).

\section{Appendix A: Derived formulas for the Eight Cases in Section II}

The two tables below summarizes the important formulas derived in the four odd cases and four even cases in Section I. 

\bgroup
\def\arraystretch{2.0}% 
\begin{table*}[t!]
\centering
%     \begin{adjustbox}{width=\textwidth,center}
    % \begin{adjustbox}{center}
        \begin{tabular}{|c|c|c|c|c|}
        \hline
        Case& $A(\times 4ke^2)$ &$y_p$ & $\Delta(y=y_p)$ & $W(s)$ \\\cline{1-5}
        A& $1/(R\theta$) &$\frac{1}{2}\gamma R\theta^2$ & $\frac{1}{6}R\theta^3$ & \multirow{6}{*}{$ A \times \lambda \left(s-\Delta(y=y_p) \right)$}  \\\cline{1-4}
        
        C& $1/(R\phi+2x)$ &$\frac{1}{2}\gamma (R\phi+2x)\phi$ & $\frac{1}{6}(R\phi+3x)\phi^2$ & \\\cline{1-4} 
        \multirow{2}{*} E& $(\phi_1 \pm \theta_2)/(R_1\phi_1^2 +2x\phi_1$ &$\frac{1}{2}\gamma (R_1\phi_1^2+2x\phi_1 $ 
        & $\frac{1}{6}[(R_1\phi_1+3x)\phi_1^2$ & \\
        
        & $+2R_2\phi_1\theta_2 \pm R_2\theta_2^2)$&$+2R_2\phi_1\theta_2 \pm R_2\theta_2^2)$ & $+R_2 \theta_2(3\phi_1^2 \pm 3\phi_1\theta_2+\theta_2^2)]$  &\\\cline{1-4}
        
        \multirow{2}{*} G& $(\phi_1 \pm \phi_2)/(R_1\phi_1^2+2x\phi_1$ &$\frac{1}{2}\gamma (R_1\phi_1^2+2x\phi_1+2R_2\phi_1\phi_2$ & $\frac{1}{6}[(R_1\phi_1+3x)\phi_1^2+R_2\phi_2(3\phi_1^2$ & \\ 
        
        & $+2R_2\phi_1\phi_2 \pm R_2\phi_2^2 +2x_2(\phi_1 \pm \phi_2))$ &$\pm R_2\phi_2^2 + 2x_2(\phi_1 \pm \phi_2))$ & $\pm 3\phi_1\phi_2+\phi_2^2)+3x_2(\phi_1 \pm \phi_2)^2]$  & \\\hline
        \end{tabular}
    \caption{ The normalization factor $A$, the $y_p$ which approximately maximizes $w(y)$, and the wake expression $W(s)$ for odd cases. The $``\pm"$ sign indicates $``+"$ if the two magnets bend in the same direction, and $``-"$ if opposite. Note that the expressions for $W(s)$ can only be applied if the drift length before the (first) magnet is much longer than $y_p$.}
    \label{tb:odd}
\end{table*}
\egroup

\bgroup
\def\arraystretch{2.0}% 
\begin{table*}[t!]
\centering
        \begin{tabular}{|c|c|c|c|}
        \hline
        Case& $u(\times 4ke^2)$ & Boundary term(s) of $W(s)$ &  Integral term of $W(s)$\\\cline{1-4}
        B& $1/(R\theta)$& $-u(\theta=\theta_{\text{max}})\lambda(s-\Delta_{\text{max}})$& \multirow{4}{*}{\( \displaystyle - \int^{s-\Delta_{\text{min}}}_{s-\Delta_{\text{max}}} u\frac{\partial \lambda(s')}{\partial s'} ds'\)}\\\cline{1-3}
        D& $1/(R\theta+2x)$& $-u(\theta=\phi)\lambda(s-\Delta_{\text{max}})$ & \\\cline{1-3}
        F& \(\displaystyle
\frac{(\theta_1 \pm \theta_2)}{R_1\theta_1^2 +2x\theta_1 \pm 2R_2\theta_1\theta_2 \pm R_2\theta_2^2}\) & \multirow{2}{*}{$u(\theta_1=0)\lambda(s-\Delta_{\text{min}})$}&\\\cline{1-2}
        H& \(\displaystyle \frac{(\theta_1 \pm \phi_2)}{R_1\theta_1^2 +2x\theta_1 \pm 2R_2\theta_1\phi_2 \pm R_2\phi_2^2+2x_2(\theta_1 \pm \phi_2)} \)&$-u(\theta_1=\phi_1)\lambda(s-\Delta_{\text{max}})$& \\\hline
        \end{tabular}
    \caption{The function $u$ and the wake expression $W(s)$ for even cases. Note that for case B and D there is one boundary term, and for case F and H there are two boundary terms (the near one at $\theta_1=0$ and  the far one at $\theta_1=\phi_1$).}
    \label{tb:even}
\end{table*}
\egroup

\end{document}